\documentclass[10pt,journal,cspaper,compsoc]{IEEEtran}
\usepackage{ifpdf}

\ifpdf
\else
\fi
\ifCLASSOPTIONcompsoc
   \usepackage[nocompress]{cite}
\else
   \usepackage{cite}
\fi
\ifCLASSINFOpdf
   \usepackage[pdftex]{graphicx}
  \usepackage{graphics}
   \graphicspath{{images/}{charts/}}
   \DeclareGraphicsExtensions{.pdf,.jpg,.png}
\else
   \usepackage[dvips]{graphicx}
  \graphicspath{{images/}{charts/}}
   \DeclareGraphicsExtensions{.eps}
\fi
\usepackage[cmex10]{amsmath}
\usepackage{amsfonts}
\usepackage{array}
\usepackage{mdwmath}
\usepackage{mdwtab}
\usepackage{graphicx}
\usepackage{eqparbox}
\ifCLASSOPTIONcompsoc
  \usepackage[caption=false,font=normalsize,labelfont=sf,textfont=sf]{subfig}
\else
  \usepackage[caption=false,font=footnotesize]{subfig}
\fi
\usepackage{url}
\usepackage{xspace}

\newcommand{\ff}{\mathbb{F}}
\newcommand{\zz}{\mathbb{Z}}
\newcommand{\ts}{\tilde{S}}
\newcommand{\tsk}{\tilde{SK}}
\renewcommand{\aa}{\mathbb{A}}
\renewcommand{\gg}{\mathbb{G}}
\newcommand{\ba}{\textbf{A}\xspace}
\newcommand{\bb}{\textbf{B}\xspace}
\newcommand{\bc}{\textbf{C}\xspace}
\newcommand{\rabe}{PIRATTE\xspace}
\newcommand{\easier}{EASiER\xspace}

\newcommand{\squishlist}{
 \begin{list}{$\bullet$}
	{ \setlength{\itemsep}{0pt}
    \setlength{\parsep}{3pt}
    \setlength{\topsep}{3pt}
    \setlength{\partopsep}{0pt}
    \setlength{\leftmargin}{1.0em}
    \setlength{\labelwidth}{1em}
    \setlength{\labelsep}{0.5em}} }
\newcommand{\squishend}{\end{list}}

\begin{document}
\title{PIRATTE: Proxy-based Immediate Revocation of ATTribute-based Encryption}
\author{Sonia~Jahid and Nikita~Borisov\\
\texttt{\{sjahid2,nikita\}@illinois.edu}\\
University of Illinois at Urbana-Champaign
}


\IEEEcompsoctitleabstractindextext{%
\begin{abstract}
Access control to data in traditional enterprises is typically enforced through reference monitors. However, as more and more enterprise data is outsourced, trusting third party storage servers is getting challenging. As a result, cryptography, specifically Attribute-based encryption (ABE) is getting popular for its expressiveness. The challenge of ABE is revocation.

To address this challenge, we propose \rabe, an architecture that supports fine-grained access control policies and dynamic group membership. \rabe is built using attribute-based encryption; a key and novel feature of our architecture, however, is that it is possible to remove access from a user without issuing new keys to other users \emph{or} re-encrypting existing ciphertexts.  We achieve this by introducing a proxy that participates in the decryption process and enforces revocation constraints. The proxy is minimally trusted and cannot decrypt ciphertexts or provide access to previously revoked users.  We describe the \rabe construction and provide a security analysis along with performance evaluation. We also describe an architecture for online social network that can use \rabe, and prototype application of \rabe on Facebook.
\end{abstract}

}

\maketitle

\IEEEdisplaynotcompsoctitleabstractindextext

\section{Introduction}\label{intro}

Access control to data in traditional enterprises is typically provided by reference monitors that enforce a particular policy. This approach, however, creates a vulnerability for monitors that are buggy or compromised and thus do not enforce the correct policy. This is a particular concern in large, distributed enterprises, as well as Online Social Networks (OSNs), where recent privacy compromises showcase the dangers of entrusting privacy controls to social network providers~\cite{facebook-email-snafu,fb-article}. The problem gets challenging as enterprises outsource more and more data, and rely on third party storage servers to enforce the access policy.

Attribute-based encryption (ABE)~\cite{fuzzyibe,cpabe,kpabe} provides an alternative approach to data protection, where the ability to decrypt data items is controlled by a policy specified in terms of attributes. ABE systems mimic the expressiveness of traditional access control systems, but use cryptography instead of reference monitors. In enterprises, this means that data can be sent over channels and stored on media that might at some point become compromised without fear of a privacy breach. Among several existing schemes, Ciphertext Policy ABE (CP-ABE)~\cite{cpabe} is appropriate for most applications. In CP-ABE, the encryptor uses some public parameters and a policy described over attributes to encrypt a piece of data. Different secret keys are issued for different sets of attributes. A key that has enough attributes to satisfy the policy decrypts the ciphertext. 

Various applications can benefit from the flexibility and expressiveness of ABE, but require the support of frequent revocation. ABE falls short in such areas when frequent and immediate revocation of access is required. Researchers have proposed revocation by attaching an expiry date to the keys~\cite{cpabe,kpabe} or introducing proxies~\cite{yu:asiaccs10}. However, existing approaches come with their shortcomings either by introducing delay in revocation, increasing the size of ciphertext, or affecting (re-keying) all the users including both the revoked and non-revoked ones.

In this paper, we present \rabe, a proxy-based immediate revocation scheme for CP-ABE. Our design makes use of a minimally trusted proxy, which handles revoked users and attributes. Upon revocation, no new key is generated for any user, neither is the existing data re-encrypted. We believe this feature is key for access control in any context where ABE is used together with highly dynamic group membership and large datasets.  Note that the proxy is minimally trusted: it cannot decrypt by itself, and even if it were compromised, it cannot allow previously revoked users to decrypt either. The only assumption we hold is that the proxy is updated with a new key each time a revocation takes place. \rabe ensures forward secrecy, backward secrecy with some assumptions, immediate revocation of complete or partial access, and delegation of access with single and multiple key authorities.

\subsection*{Our Contribution:}
\squishlist
\item We provide the construction of our scheme including Proxy-based key/attribute revocation and access delegation, and the security analysis of the scheme.
\item We implement a prototype named \rabe and compare its performance with the CP-ABE scheme by Bethercourt et al.~\cite{cpabe} (BSW CP-ABE).
\item In addition, we describe a case study that can benefit from using \rabe. We choose OSN since recent research in OSNs proposes the use of cryptography to enhance privacy~\cite{persona,guha:wosn08,lucas:wpes08}. However, they are not completely successful because of some shortcomings of existing cryptographic schemes. We also present an application of \rabe running on Facebook platform.
\squishend

\subsection*{Roadmap:}
The rest of the paper is organized as follows. We briefly discuss background information on CP-ABE and the base revocation scheme in Section~\ref{background}. Next, we provide a detailed description of our construction in Section~\ref{cons}. Section~\ref{sec} discusses the security of our construction. We describe the applicability of \rabe to OSNs in Section~\ref{arch}. We describe our performance analysis and the Facebook application in Section~\ref{exp}, related work in Section~\ref{sec:related}, and conclude in Section~\ref{conclude}.

\section{Background} \label{background}

In this section we describe some background information necessary to understand our scheme. We describe two cryptographic schemes that form the foundation of our approach.

\subsection{Attribute-Based Encryption}
Cryptography can be used to enforce access control to information by encrypting data such that only authorized users can decrypt it.  However, with standard public-key cryptography, it is necessary to explicitly enumerate all of the users who may decrypt each data item.  While it is possible to issue group keys, this creates complex key management issues and several problems still remain.

Attribute-based encryption provides a better solution for defining fine-grained access control to groups of people. With ABE, a user Alice assigns sets of attributes to other users (we will call these users Alice's contacts) and issues the corresponding secret keys. She encrypts data items using policies expressed in terms of plain attributes that defines what attributes one must have in order to decrypt the piece of data (this variant is called \emph{ciphertext-policy} ABE; \emph{key-policy} ABE reverses the process, with attributes assigned to data items and policies to keys).

ABE can support complex policies, such as ``\textit{(Friend OR Co-worker) AND Neighbor}''. ABE also explicitly prevents collusion between users: if Bob is Alice's co-worker, and Carol is her neighbor, they cannot combine their attributes together to satisfy the above policy if neither of them satisfies it individually. Finally, ABE provides public-key functionality, allowing, for example, Bob to encrypt a message with Alice's public key to be decrypted by her contacts to whom Alice assigns secret attribute keys.

We can formally define a CP-ABE scheme by four algorithms with an option for attribute delegation~\cite{cpabe}:

\squishlist
\item \textbf{\textsc{Setup}}. This algorithm takes security parameters and generates a public key $PK$ and master secret key $MK$.

\item \textbf{\textsc{Encrypt$(PK,M,P)$}}. This algorithm takes the public key $PK$, a message $M$, and a policy $P$ and generates a ciphertext $CT$ encrypted with $P$.

\item \textbf{\textsc{Keygen}$(MK,S)$}. This algorithm uses the master secret key $MK$ to generate a secret attribute key $SK$ using the attributes in the set $S$.

\item \textbf{\textsc{Decrypt}$(CT,SK)$}.  This algorithm decrypts a ciphertext $CT$ to plaintext $M$ as long as the set of attributes $S$ in $SK$ satisfies the policy $P$ that was used to generate $CT$ from $M$.  (The policy is implicitly encoded in $CT$.)

\item \textbf{\textsc{Delegate}$(SK, \ts)$}.  The delegate algorithm takes as input a secret key $SK$ for some set of attributes $S$ and a set $\ts \subseteq S$. It outputs a secret key $\tsk$ for the set of attributes $\ts$.
\squishend
\subsection{Revocation Scheme}
To support practical revocation in \rabe, we adapt the broadcast revocation scheme developed by Naor and Pinkas to prevent digital piracy~\cite{naor}. The scheme uses Shamir secret sharing~\cite{shamirsecret} to create shares of a secret key, where $t+1$ shares are necessary for reconstruction, and gives one share to each user.

During regular operation, the distributor broadcasts $t$ random shares, which lets any user to reconstruct the secret key by combining the shares with his or her own. To revoke up to $t$ users, the distributor broadcasts their shares instead of the random ones. Any non-revoked user still has $t+1$ distinct shares and can reconstruct the secret, whereas the revoked users do not have enough information even if they all collude.
\section{Construction}\label{cons}

\subsection{Assumptions and Basics} \label{basics}
Before going into the details of the construction, we present some basic
mathematical assumptions, and the details of CP-ABE and the revocation scheme
used in \rabe.

\subsection*{Bilinear Pairing} Let $\gg_{1}$, $\gg_{2}$, and
$\gg_{T}$ be multiplicative cyclic groups of prime order $p$, and $e$ a map
($\gg_{1} \times \gg_{2} \rightarrow \gg_{T}$).  Let  $g_{1}$ and $g_{2}$ be
generators of $\gg_{1}$ and $\gg_{2}$ respectively ($\gg_i=\langle g_i
\rangle$).  If $\forall  u \in \gg_{1}, v \in \gg_{2}$ and $ a, b \in \zz_p $,
$e(u^{a},v^{b}) = e(u,v)^{ab}$ and $e(g_{1}, g_{2}) \neq 1$, then $e$ is called
a bilinear pairing. If $\gg_{1} = \gg_{2}$, it is called a symmetric pairing,
otherwise the pairing is asymmetric.


\subsection*{Secret Sharing} In Shamir's secret sharing
scheme~\cite{shamirsecret}, a secret $s$ in some field $F$ is shared among $n$
parties by creating a random polynomial $P \in F[x]$ of degree $t$ such that $P(0) = s$.
 The $i$-th party gets the share $\langle i,P(i) \rangle$.  Given any $t+1$ shares
$P(x_0),\ldots,P(x_t)$, it is possible to recover $P(0)$ using Lagrange
interpolation:
\begin{align*}
& P(0) = \sum_{i=0}^{t}\lambda_{i}P(x_{i}), \quad \text{where }\lambda_{i} =
\prod_{j \neq i} \frac{x_{j}}{(x_{j}-x_{i})}
\end{align*}

\subsection*{BSW CP-ABE Scheme} The algorithms in CP-ABE due to the Bethencourt et al. are described below.
Though CP-ABE uses symmetric pairing, it can be implemented using an asymmetric
pairing as well.

\squishlist
\item \textbf{\textsc{Setup:}} The key authority KA generates a public key PK, and a master
secret key MK:
\begin{align*}
& PK = \gg_{1}, g, h = g^{\beta},  e(g,g)^{\alpha}\\
& MK = (\beta,g^{\alpha}) \\
& \text{where random } \alpha, \beta \in \zz_{p}, \gg_1=\langle g \rangle , |\gg_1 | = p
\end{align*}
The $PK$ also contains an extra component $f=g^{1/\beta}$ to support attribute delegation.

\item \textbf{\textsc{Encrypt(PK, M, $\tau$):}} A policy is represented as an access tree
structure $\tau$ with the attributes at leaves and threshold $k$-of-$n$ gates 
at intermediate nodes. Each node is associated with a polynomial $q_{x}$ of
degree $ d_{x} $, where $ d_{x} $ is $1$ less than the threshold value $k$ of
that node. The polynomials are of degree $0$ for OR gates and leaves. The secret
$s$ (random $ s \in \zz_p $) to blind the data $M$ is associated with the
polynomial at the root of the tree, i.e., $q_{R}(0) = s$. The sharing works in a top
down manner: for all other nodes, $q_{x}(0) = q_{parent(x)}(index(x))$. 
$index(x)$ returns a number between $1$ and $num$ associated with $x$ where
$num$ is the number of children of $parent(x)$. Let $Y$ be the set of leaf nodes
in $\tau$. The ciphertext $CT$ is:
\begin{align*}
CT & = (\tau , \tilde{C} = Me(g,g)^{\alpha s}, C=  h^{s},\\
& \forall y \in Y: C_{y} = g^{q_{y}(0)}, C'_{y} = H(att(y))^{q_{y}(0)})
\end{align*}
Here, $H : \{0,1\}^{*} \rightarrow \gg_{1}$ is a hash function, modeled as
random oracle, that maps string attribute to random element of $\gg_{1}$.

\item \textbf{\textsc{KeyGen(MK, S):}} The secret key $SK$ corresponding to a set of
attributes $S$ is (random  $r, r_{j} \in \zz_p$):
\begin{align*}
 SK =&  (D = g^{(\alpha + r)/\beta}, \\
  & \forall j \in S : D_{j} = g^{r} H(j)^{r_{j}}, D'_{j} = g^{r_{j}})
\end{align*}

$D_{j}, D_{j}'$ for each attribute are blinded by $r_{j}$, and all the
components are tied together using $r$ in $D$. This prevents attributes of
different users from being combined together and provides collusion resistance.

\item \textbf{\textsc{Decrypt(CT, SK):}} The goal of decryption algorithm is to find out
$e(g,g)^{\alpha s}$. It finds out the secret $q_{x}(0)$ at each node $x$ blinded
by the random value $r$. A secret key $SK$ that achieves $d_{R}$ such secrets at the root
$R$, can solve the polynomial $q_{R}$ and decrypt the ciphertext. A recursive
algorithm $DecryptNode$ pairs $D_{i}$ and $D_{i}'$ (from $SK$) with $C_{x}$ and
$C_{x}'$ (from $CT$) respectively and returns $e(g,g)^{rq_{x}(0)}$ for each leaf
node $x$ in the $\tau$ in $CT$, iff $i = attr(x)$. $ i \in S$ is the set of
attributes for which a user is assigned $SK$. 

At each non-leaf node, Lagrange
interpolation is used on at least $k$ (the threshold value of the node) such
$e(g,g)^{rq_z{0}}$ received from its children $z$, to calculate
$e(g,g)^{rq_{x}(0)}$. Let $A = e(g,g)^{rq_{R}(0)} = e(g,g)^{rs}$. Then
$\tilde{C}, C, D$ and $A$ are used in bilinear mapping to cancel out
$e(g,g)^{rs}$, and retrieve $M$. Further details can be found in~\cite{cpabe}.

\item \textbf{\textsc{Delegate}$(SK, \ts)$}. The delegate algorithm re-randomizes the relevant set of attributes $\ts \subseteq S$ of a secret key $SK$ assigned for some set of attributes $S$. It outputs a secret key $\tsk$ for the set of attributes $\ts$.
\vspace{-10pt}
\begin{align*}
\tsk =	& (\tilde{D}=Df^{\tilde{r}}, \\
	& \forall k \in \ts: \tilde{D_k} = D_k g^{\tilde{r}} H(k)^{\tilde{r}_k}, \tilde{D}'_k = D'_k g^{\tilde{r}_k})
\end{align*}
\squishend
\vspace{-20pt}
\subsection*{Revocation Scheme of Naor and Pinkas} This scheme
consists of $2$ phases:

\squishlist
\item \textbf{Initialization:} The group controller  generates  a random polynomial
$P$ of degree $t$ over $\zz_p$. It sends a personal key $\langle I_u, P(I_u)
\rangle$ to each user $u$ with random identity $I_u$. This process is  performed only
once for all future revocations.

\item \textbf{Revocation:} The group controller learns the random identities of $t$ users
$I_{u_1},\ldots,I_{u_t}$ that should be revoked. At most $t$ users can be revoked since the scheme depends on polynomial secret sharing. It then chooses a random $r$, and
sets the new key to be $g^{rP(0)}$, that would be unknown to revoked users. It
broadcasts the message $g^r, \langle I_{u_1}, g^{rP(I_{u_1})} \rangle, \ldots.,
\langle I_{u_t}, g^{rP(I_{u_t})} \rangle$ encrypted with the current group key.
Each non revoked user can compute $g^{rP(I_u)}$ and combine it with the
broadcast values to obtain $g^{rP(0)}$ using Lagrange's interpolation formula.
Further details can be found in~\cite{naor}.
\squishend
\subsection{Proxy-based Complete Key Revocation}
In this section we describe how to completely revoke keys from parties. That
means, all the privileges granted by the key authority are revoked from one or more
contact(s). This construction allows revocation of up to $t$ users at a time
since it is based on the scheme in~\cite{naor} described before.

\subsection*{Intuition:}
The master key $MK$ contains a polynomial $P$ of degree $t$. $P(0)$ is used to blind users' secret keys. Each user $u$ also gets a random share $P(u)$ of $P(0)$ in her key. The proxy key consists of $t$ such shares and is used to convert a part of the ciphertext for decryption. Whenever access is revoked from someone, her share becomes a part of the proxy key, and eventually the converted ciphertext. Therefore, the revoked user does not have enough points, i.e. $(t+1)$ points to unblind her key and the ciphertext and decrypt it. However, non-revoked users can always combine their secret keys with the ciphertext and hence decrypt it. 

When no one is revoked, the proxy key consists of $t$ random $P(u)$ points. Since the revocation is based on polynomial secret sharing, and the degree of the polynomial is $t$, the scheme is limited to maximum $t$ revocations. Though each time $t$ different users can be revoked, the total number of users in the system is not limited.

\textbf{\textsc{- Setup:}} The key authority KA randomly generates a polynomial $P$ of
degree $t$ (the maximum number of revoked users) over $\zz_p$, sets the
broadcast secret $P(0)$ to be used after revocation, and randomly chooses
$\alpha, \beta \in \zz_p$. She generates $PK$ and $MK$ as follows:
\begin{align*}
& PK = \gg_{1}, \gg_{2}, g_1, g_2, h = g_{1}^{\beta},  e(g_1,g_2)^{\alpha}\\
& MK = \beta,g_{2}^{\alpha}, P
\end{align*}

\textbf{\textsc{- Encrypt(PK, M, $\tau$):}} Let $Y$ be the set of leaf nodes in $\tau$.
Data $M$ is encrypted to get the ciphertext $CT$. Other than the asymmetric groups, this algorithm works exactly
the same as in BSW CP-ABE.
\begin{align*}
CT & = (\tau , \tilde{C} = Me(g_1,g_2)^{\alpha s}, C = h^{s} = g_1^{\beta s},\\
& \forall y \in Y: C_{y} = g_1^{q_{y}(0)},\  C'_{y} = H(att(y))^{q_{y}(0)} =
g_2^{h_{y}q_{y}(0)} )
\end{align*}
where $H : \{0, 1\}^* \rightarrow \gg_2$ and $h_y = \log_{g_2} H(att(y))$ (used for
notational convenience only).

\textbf{\textsc{- KeyGen(MK, S):}} The algorithm KeyGen outputs the secret key
corresponding to the set of attributes $S$, blinded by $P(0)$ from $MK$. We introduce an extra component--- $D_{j}''$---that in addition to attribute information contains user information. Without loss of
generality, we assume user $u_{k}$ receives this key.
\begin{align*}
SK  &= (D, \forall j \in S:  \langle D_{j}, D_{j}', D_{j}'' \rangle),
\text{where} \\
& D = g_2^{(\alpha + r)/\beta}, \\
& D_{j} = g_2^{r} H(j)^{r_{j}P(0)} = g_2^{r+h_{j}r_{j}P(0)}, \\
& D'_{j} = g_1^{r_{j}},\\
& D''_{j} = (D'_{j})^{P(u_{k})} = g_1^{r_{j}P(u_{k})}
\end{align*}

\textbf{\textsc{- ProxyRekey(PK, MK, $RL$):}} Whenever the KA wants to revoke keys
from social contacts, she creates a list of revoked users $RL$ with their
identities $u_{i}$, $i \in \{1, \ldots, t\}$, and evaluates the corresponding
$P(u_{i})$ using $MK$. She gives the proxy key $PXK$ to the proxy. In case of no
or fewer than $t$ revocations, the KA generates random $\langle x,P(x)
\rangle$ other than the actual user identities, to make $PXK$ of length $t$.
\begin{align*}
& PXK = \forall u_i \in RL:  \langle u_{i},P(u_{i}) \rangle
\end{align*}

\textbf{\textsc{- Convert(PXK, $\forall y \in Y: C_{y} $, $u_{k}$):}} The proxy uses its
key $PXK$ and the decryptor's identity $u_{k}$ to calculate $C_{y}''$ as follows:
\begin{align*}
& \forall i,j \in \{1, \ldots, t\},\ k \notin \{1, \ldots, t\}, \\
& \lambda_{i} = \frac{u_k}{u_k-u_i}\cdot \prod_{j \neq i}
\frac{u_{j}}{(u_{j}-u_{i})}, \\
& \forall y \in Y: C_{y}'' = (C_{y}')^{\sum_{i=1}^{t} \lambda_{i}P(u_i)} =
g_2^{h_{y}q_{y}(0)\sum_{i=1}^{t} \lambda_{i}P(u_{i})}
\end{align*}
Since the user secret key $SK$ is blinded by $P(0)$, she needs $C_{y}''$ in
addition to $C_{y}$ and $C_{y}'$ for decryption.The proxy also calculates
$\lambda_{k}$ and gives it to the user $u_{k}$.

\textbf{\textsc{- Decrypt(CT, SK):}} The decryption steps involve one extra pairing than
BSW CP-ABE at each leaf node of the policy. For each leaf node $x$ where $i =
attr(x)$, if $i \in S$, ($S$ is the set of attributes for which $SK$ is issued)
then,
\begin{align*}
   & DecryptNode (CT, SK, x) \\
   & = \frac{e(C_{x},D_{i})}{e(D''_{i}, C'_{x})^{\lambda_{k}} e(D'_{i},
C''_{x})} \\
   & =
\frac{e(g_{1},g_{2})^{rq_{x}(0)+h_{i}r_{i}P(0)q_{x}(0)}}{e(g_{1},g_{2})^{r_{i}h_
{i}q_{x}(0)\lambda_{k}P(u_{k})} e(g_{1},g_{2})^{r_{i}h_{i}q_{x}(0)\sum_{j=1}^{t}
\lambda_{j}P(u_{j}) }}\\
   & =
\frac{e(g_{1},g_{2})^{rq_{x}(0)+h_{i}r_{i}P(0)q_{x}(0)}}{e(g_{1},g_{2})^{r_{i}h_
{i}q_{x}(0) (\sum_{j=1}^{t} \lambda_{j}P(u_{j}) + \lambda_{k}P(u_{k}))}} \\
   & =
\frac{e(g_{1},g_{2})^{rq_{x}(0)+h_{i}r_{i}P(0)q_{x}(0)}}{e(g_{1},g_{2})^{r_{i}h_
{i}q_{x}(0)P(0)}} , k\not \in \{1, 2, \ldots, t\}\\
   & = e(g_{1},g_{2})^{rq_{x}(0)}
\end{align*}
Otherwise $DecryptNode$ returns $\bot$. The rest of the decryption is the same
as CP-ABE. For each child $z$ of a non-leaf node x, it calculates $F_z =
e(g_{1},g_{2})^{rq_z(0)}$. Let $S_{x}$ be a threshold-sized arbitrary set of
children of x, such that $F_z \neq \bot$. Then interpolation and pairings are
used to calculate $e(g_{1},g_{2})^{\alpha s}$, and hence retrieve $M$.
\begin{align*}
F_x &= \prod_{z \in S_{x}}  F_z^{\lambda_{i}}, [i=index(z),\\
& \lambda_i \text{ calculated over the indices of }  z \in S_x]\\
&= \prod_{z \in S_{x}} (e(g_1, g_2)^{rq_z(0)})^{\lambda_{i}}\\
&= \prod_{z \in S_{x}} (e(g_1, g_2)^{rq_{parent(z)}(index(z))})^{\lambda_{i}},
[\text{discussed in~\ref{basics}}]\\
&= \prod_{z \in S_{x}} (e(g_1, g_2)^{rq_{x}(i)})^{\lambda_{i}}\\
&= e(g_1, g_2)^{\sum_{z \in S_{x}}{r \lambda_i q_x(i)}}\\
&= e(g_1, g_2)^{rq_x(0)}
\end{align*}
Let $A=e(g_1,g_2)^{rq_R(0)} = e(g_1,g_2)^{rs}$ at the root $R$. Decryption proceeds as follows,
\begin{align*}
\frac{\tilde{C}}{\frac{e(C,D)}{A}} =
M e(g_1,g_2)^{\alpha s} \frac{e(g_1,g_2)^{rs}}{e(g_1,g_2)^{\alpha s + rs}} = M
\end{align*}
\vspace{-18pt}
\subsection*{Explanation of Asymmetric Group:}
We use different groups for $C'_i$ and $D'_i$ ($i$ is the attribute). The user gets $C'_i$ converted to $C''_i = {C'_i}^ a $ ($a = {\sum_{j=1}^{t} \lambda_{j}P(u_j)}$, explained in the \emph{Convert} algorithm) by the proxy. However, if both $C'_i$ and $D'_i$ belong to the same group, and the user gives the proxy $D'_i$ instead of $C'_i$, she will get $ (D'_i)^a = g^{a r_i}$.  She will also get $\lambda_k$, which she can use to get ${D''_i}^{\lambda_k} = g^{\lambda_k P(u_k) r_i}$. Multiplying these two, she gets $g^{r_i  (a + \lambda_k P(u_k))} = g^{r_i P(0)} $.  She can use this last value to decrypt any ciphertext without using the proxy, so the revocation is no longer effective. Therefore, we use asymmetric pairing, where $C'_i$ and $D'_i$ are in different groups and mapping of $D'_i$ into the $C'_i$ group is not possible.

\vspace{-18pt}
\subsection{Delegation of Access} \label{delegate}

We design delegation of attributes in \rabe in two settings - 1) When all the keys to different parties are issued by a single key authority, and 2) When keys are issued by multiple key authorities.

\subsubsection{Single Key Authority}
In the single authority setting, a user $u_k$ gets a secret key $SK$ from key authority KA, and delegates one or more of the attributes that she possesses in her secret key to another user. As long as $u_k$ is not revoked, the delegated key can be used for decryption. The delegation process is as follows. The delegation algorithm takes in a secret key $SK$ issued for a set of attributes $S$ and delegates one or more attributes from this set. The delegated key $\tsk$ is generated for the subset of attributes $\ts \subseteq S$. For delegation in single authority setting, an extra public parameter $f=g_2^{1/\beta}$ is introduced in the public key $PK$. Let random $\tilde{r} \in Z_p$, and random $\forall_j \in \ts, \tilde{r}_j \in Z_p$.
\begin{align*}
\tilde{SK} &=(\tilde{D},{\forall}j\in\ts: \langle \tilde{D}_j,D_j', \tilde{D}_j''{\rangle}), \text{where}\\
& \tilde{D} = (D)f^{\tilde{r}} = g_2^{(\alpha + r + \tilde{r})/\beta}\\
& \tilde{D}_j = (D_j)g_2^{\tilde{r}} H(j)^{\tilde{r}_j} = g_2^{r+\tilde{r} + h_j r_j P(0) + h_j \tilde{r}_j }\\
& \tilde{D}''_j = (D''_j) g_1^{\tilde{r_j}/\lambda_{k}} = g_1^{r_j P(u_k) + \tilde{r}_j/\lambda_k}
\end{align*}

\noindent Decryption proceeds are follows: 
\begin{align*}
   & DecryptNode (CT, \tsk, x) \\
   & = \frac{e(C_{x},\tilde{D}_{i})}{e(\tilde{D}''_{i}, C'_{x})^{\lambda_{k}} e(D'_{i},
C''_{x})} \\
   & =
\frac{e(g_{1},g_{2})^{q_x(0)(r+\tilde{r} + h_i r_i P(0) + h_i\tilde{r}_i)}}{e(g_{1},g_{2})^{(r_i P(u_{k}) + \tilde{r_i}/\lambda_k) h_i q_x(0) \lambda_k}}  \cdot\\
 & \frac{1}{e(g_{1},g_{2})^{r_{i}h_{i}q_{x}(0)\sum_{j=1}^{t} 
\lambda_{j}P(u_{j}) }}\\
   & =
\frac{e(g_{1},g_{2})^{(r+\tilde{r})q_{x}(0) + h_{i}r_{i}P(0)q_{x}(0) + h_{i}\tilde{r}_{i}q_{x}(0)}}
{e(g_{1},g_{2})^{r_{i}h_{i}q_{x}(0)\lambda_{k}P(u_{k}) + h_i \tilde{r}_i q_x(0)} } \cdot \\
& \frac{1}{e(g_{1},g_{2})^{r_{i}h_{i}q_{x}(0)\sum_{j=1}^{t}
\lambda_{j}P(u_{j}) }}\\
   & =
\frac{e(g_{1},g_{2})^{(r+\tilde{r})q_{x}(0) + h_{i}r_{i}P(0)q_{x}(0) + h_{i}\tilde{r}_{i}q_{x}(0)}}
{e(g_{1},g_{2})^{r_{i}h_{i}q_{x}(0) (\sum_{j=1}^{t} \lambda_{j}P(u_{j}) + \lambda_{k}P(u_{k})) + h_i \tilde{r}_i q_x(0)}} \\
   & =
\frac{e(g_{1},g_{2})^{(r+\tilde{r})q_{x}(0)+h_{i}r_{i}P(0)q_{x}(0)+ h_i \tilde{r}_i q_x(0)}}
{e(g_{1},g_{2})^{r_{i}h_{i}q_{x}(0)P(0) + h_i \tilde{r}_i q_x(0)}} , \\ & k\not \in \{1, 2, \ldots, t\}\\
   & = e(g_{1},g_{2})^{(r + \tilde{r})q_{x}(0)}
\end{align*}

Let $A=e(g_1,g_2)^{(r+\tilde{r})q_R(0)} = e(g_1,g_2)^{(r+\tilde{r})s}$ for the root node. The rest of the decryption proceeds as follows,
\begin{align*}
\frac{\tilde{C}}{\frac{e(C,\tilde{D})}{A}} =
M e(g_1,g_2)^{\alpha s} \frac{e(g_1,g_2)^{(r+\tilde{r})s}}{e(g_1,g_2)^{\alpha s + rs + \tilde{r}s}} = M
\end{align*}

\subsubsection{Multiple Key Authority}

The second version of delegation of access is designed with a distributed setting in mind, i.e.,
when secret attribute keys are issued from different key authorities to
different users. In this setting, \ba generates keys for \bb and \bb generates keys for \bc; i.e., \bb is a contact of \ba and \bc is a contact of \bb. Again, the delegation algorithm takes in a secret key $SK$ issued for a
set of attributes $S$ and delegates one or more attributes from this set. In the following construction, we will show how \bb delegates a key $SK$ for the set of attributes $S$ generated by \ba for \bb, to \bc.
\begin{align*}
SK  &= (D, \forall j \in S:  \langle D_{j}, D_{j}', D_{j}'' \rangle),
\text{where}\\
& D = g_{2}^{(\alpha + r)/\beta}, \\
& D_{j} = g_{2}^{r} H(j)^{r_{j}P_A(0)}, D'_{j} = g_{1}^{r_{j}}, D''_{j}
= (D'_{j})^{P_A(B)}
\end{align*}
where random  $r,r_j \in Z_p$, $P_A$ is the polynomial in \ba 's $MK$, and $B$ is
\bb's identity.

Attribute delegation allows \bb to delegate some subset of attributes $\ts
\subseteq S$ to his contact \bc. \bb re-randomizes $SK$ for \bc for a set of
attributes $\ts \subseteq S$. 
\begin{align*}
\tilde{SK} &=(D,{\forall}j{\in}\ts:{\langle}D_j,D_j',
\tilde{D_j''}, \tilde{D_j'''}{\rangle}),\text{where}\\
& \tilde{D''_j}=(D''_j)^{1/P_B(0)}, \tilde{D'''_j}=(D''_j)^{P_B(C)/P_B(0)}
\end{align*}
where $P_B$ is the polynomial in \bb's $MK$, and $C$ is \bc's identity.

To decrypt a ciphertext $CT$ encrypted with \ba's public parameters, \ba's proxy
calculates $C_{yA}'' = (C_y')^{X_A}$ and $\lambda_B$, \bb's proxy calculates
$C_{yB}'' = (C_y')^{X_B}$ and $\lambda_C$, and gives it to \bc. Here, $C_y'$ is
from $CT$, and $X_A$ and $X_B$ are revocation information for \ba's proxy and
\bb's proxy respectively, calculated as $X_A = {\sum_{i=1}^{t_1}
\lambda_{i}P_A(u_i)}, X_B = {\sum_{j=1}^{t_2} \lambda_{j}P_B(u_j)}$. $u_i$ and
$u_j$ are the revoked users by \ba and \bb respectively, $\lambda_i$ and
$\lambda_j$ are the Laggrange's coefficients for corresponding revoked users,
and $t_1$ and $t_2$ are the parameters for the maximum number of revoked users
by \ba and \bb defined in their $MK$s. $\lambda_B$ and $\lambda_C$ are the
Laggrange's coefficients for \bb and \bc respectively. \bb's identity is
conveyed to \bc as a part of the communication for $\tilde{SK}$, or any other
way. \bc does a modified bilinear pairing in the $DecryptNode$, and finally
decrypts the data.
\begin{align*}
& DecryptNode(CT,\tilde{SK},x)\\
&={\frac{e(C_x,D_i)}{e(\tilde{D''_i},C_{xB}'')^{\lambda_B}e(\tilde{D'''_i},
C_x')^{{\lambda}_B{\lambda}_C}e(D_i',C_{xA}'')}}\\
&={\frac{e(C_x,D_i)}{e(g_0,g_1)^{r_ih_xq_x(0)X_B\lambda_B P_A(B)/P_B(0)}}}\\
& . {\frac{1}{e(g_0,g_1)^{r_ih_xq_x(0)P_A(B)P_B(C)/P_B(0)\lambda_B\lambda_C}
e(D'_i,C''_{xA})}}\\
&={\frac{e(g_0,g_1)^{rq_x(0) + r_i h_x q_x(0)
P_A(0)}}{e(g_0,g_1)^{r_ih_xq_x(0)\lambda_B
P_A(B)}e(g_0,g_1)^{r_ih_xq_x(0)X_A}}}\\
&={\frac{e(g_0,g_1)^{rq_x(0) + r_i h_x q_x(0) P_A(0)}}{e(g_0,g_1)^{r_i h_x
q_x(0) P_A(0)}}}\\ &=e(g_0,g_1)^{rq_x(0)}
\end{align*}
The rest of the decryption proceeds as before. If \ba revokes \bb or \bb
revokes \bc, the decryption will not succeed since both the proxies participate
in the decryption, and the delegated secret key $\tilde{SK}$ contains
information about all \ba, \bb, and \bc.

\subsection{Proxy-based Attribute Revocation}
In this section, we describe how to revoke one or more attributes from a given
secret key. This is useful since often the KA may want to merely revoke a few
attributes from her contacts instead of the whole key. For instance, user \ba
might want to remove friend attribute from \bb, but \bb still remains in her
colleague group.

\subsection*{Intuition:}
The idea is basically the same as complete key revocation. The master key contains one polynomial $P_i$ of degree $t_i$ for each possible attribute $i$ that the KA can assign. Any attribute can be introduced later by introducing a new polynomial in the $MK$. $P_i(0)$  is used to blind the corresponding attribute in the secret keys. Each user $u$ also gets a random share $P_i(u)$ of $P_i(0)$ in her key. The proxy key consists of $t_i$ such shares for each attribute in the policy used in the ciphertext. Whenever some attribute is revoked from some user, that share becomes a part of the proxy key, and hence the converted ciphertext. Therefore, the revoked user does not have enough points, i.e. $(t_i+1)$ points for that specific attribute to unblind her key and the ciphertext and decrypt it. However, non-revoked users can always combine their secret keys with the ciphertext and hence decrypt it. As before, when no attribute is revoked, the proxy key consists of $t_i$ random points for each attribute $i$. 

\textbf{- Setup:}  The KA generates one polynomial $P_{y}$ randomly over $\zz_p$
for each attribute $y \in Y'$ where $Y'$ is an initial set of attributes in the
system, and sets $P_{y}(0)$ as the secret to be used to revoke the attribute. To
revoke an attribute from $t$ users at a time, the degree of the polynomials is
chosen to be $t$. New attributes can be introduced later by randomly generating
polynomials for them. Finally, she randomly chooses $\alpha, \beta \in \zz_p$.
\begin{align*}
& PK = \gg_1, \gg_2, g_1, g_2, h = g_1^{\beta},  e(g_1,g_2)^{\alpha}\\
& MK = \beta,g_{2}^{\alpha}, \forall_{y \in Y'}: P_{y}
\end{align*}

\textbf{- KeyGen(MK, S):} The components of the secret key are similar as before
except that the polynomial in each is specific to the attribute represented by
the component. Again, without loss of generality, we assume user $u_{k}$
receives this key.
\begin{align*}
SK  & = (D, \forall j \in S:  \langle D_{j}, D_{j}', D_{j}'' \rangle), \text{
where}\\
&  D = g_2^{(\alpha + r)/\beta}, D_j = g_2^{r} \cdot H(j)^{r_jP_j(0)} = g_2^{r +
h_j r_j P_j(0)}, \\
& D'_j = g_1^{r_j}, D''_j = (D'_j)^{P_j(u_{k})} = g_{1}^{r_{j}P_j(u_{k})}
\end{align*}

\textbf{- Encrypt(PK, M, $\tau$):} Encryption is similar as in complete key revocation.

\textbf{- ProxyRekey(PK, MK, $\forall y \in Y: RL_y $):} To revoke an attribute
$y \in Y$ from $t$ contacts, the KA creates a $t$-sized list $RL_y = \{ u_i \},\
i \in \{1, \ldots, t\}$ of revoked users for that attribute, and evaluates
$P_y(u_{i})$ using $MK$. In case of no or less than $t$ revocations, she
generates random $\langle x,P_y(x) \rangle$ to make $RL_y$ of length $t$. The
set of users from whom different attributes are revoked, may or may not overlap.
Without loss of generality we assume that the sets of revoked users don't
overlap. The proxy key $PXK$ is constructed as follows:
\begin{align*}
PXK & = \forall y \in Y, \forall u_i \in RL_y: \langle u_{i},P_y(u_{i}) \rangle
\end{align*}

\textbf{- Convert(PXK, $\forall y \in Y: C_{y}$):} The proxy uses its key $PXK$ to
convert the attribute relevant components $C_y'$ received from user $u_k$ to
$C_{y}''$ as follows:
\begin{align*}
\lambda^y_i = & \frac{u_k}{u_k-u_i} \cdot\prod_{j \neq i}
\frac{u_{j}}{(u_{j}-u_{i})},\\
& \forall u_i,u_j \in RL_y, u_k \notin RL_y, RL_y \in PXK
\end{align*}
\begin{align*}
\forall y \in Y: C_{y}'' = (C_{y}')^{\sum_{i=1}^{t} \lambda^y_{i}P_y(u_{i})} =
g_{2}^{h_{y}q_{y}(0)\sum_{i=1}^{t} \lambda^y_{i}P_y(u_{i})}
\end{align*}
$\forall y \in Y$ the proxy also calculates and gives $ \lambda^y_k $, to $u_k$.

\textbf{- Decrypt(CT, SK)}: For each leaf node $x$ where $i = attr(x)$, if $i
\in S$ ($S$ is the set of attributes for which $SK$ is issued), and $i$ is not
revoked from $u_{k}$ then,
\begin{align*}
   & DecryptNode (CT, SK, x)  \\
   & = \frac{e(C_{x}, D_{i})}{e(D''_{i}, C'_{x})^{\lambda^i_{k}} e(D'_{i},
C''_{x})}\\
   & =
\frac{e(g_1,g_2)^{rq_{x}(0)+h_{i}r_{i}P_{i}(0)q_{x}(0)}}{e(g_1,g_2)^{r_{i}h_{i}
q_{x}(0)\lambda^i_k P_{i}(u_{k})} e(g_1,g_2)^{r_{i}h_{i}q_{x}(0)\sum_{j=1}^{t}
\lambda^i_{j}P_{i}(u_{j}) }}\\
   & =
\frac{e(g_1,g_2)^{rq_{x}(0)+h_{i}r_{i}P_{i}(0)q_{x}(0)}}{e(g_1,g_2)^{r_{i}h_{i}
q_{x}(0) (\lambda^i_{k}P_{i}(u_{k}) + \sum_{j=1}^{t}
\lambda^i_{j}P_{i}(u_{j}))}} \\
   & =
\frac{e(g_1,g_2)^{rq_{x}(0)+h_{i}r_{i}P_{i}(0)q_{x}(0)}}{e(g_1,g_2)^{r_{i}h_{i}
q_{x}(0)P_{i}(0)}} , k\not \in \{1, 2, \ldots, t\}\\
   & = e(g_1,g_2)^{rq_{x}(0)}
\end{align*}
Otherwise $DecryptNode$ returns $\bot$. The rest of the decryption is as before.
In summary, if an attribute $i$ is revoked from user $u$, he can not do pairing
on $C_{x}''$ and $D_{i}'$. He can continue to use components related to his
other unrevoked attributes. Therefore, some of his attributes are revoked
whereas some continue to be active.

\section{Security Analysis}\label{sec}

First, we need to define the requisite security properties for CP-ABE with Proxy Revocation.  We present the definition for identity-based revocation; the definition for attribute-based revocation is analogous.  We base our definition on the security model defined by Bethencourt et al.~\cite{cpabe}, with the addition of revocation and proxy operations.  In this game, all encryptions remain secure even when the adversary compromises the proxy and obtains its key material, as long as this happens after the most recent revocation.

\noindent \textbf{Setup.}  The challenger runs the \textsc{Setup} algorithm and gives the public parameters, $PK$, to the adversary.  The challenger also runs \textsc{ProxyRekey}$(PK, MK, \emptyset)$ to generate a proxy key $PXK$.

\noindent \textbf{Phase 1.} The adversary makes repeated queries to \textsc{Keygen} to obtain keys for users $u_1, \ldots, u_{q_1}$ with sets of attributes $S_1, \ldots, S_{q_1}$.  The adversary also
interacts with the proxy by calling \textsc{Convert} with the input $(\{C_1', \ldots C_r' \}, u_k)$ for $C_i' \in \mathbb{G}_1$ and $u_k \in \mathbb{Z}_p$, at which point the challenger
runs the \textsc{Convert} algorithm with the stored proxy key $PXK$.  Finally, the adversary may call \textsc{ProxyRekey} by supplying a revocation list $RL$.  This will cause the challenger to update the proxy key $PXK$.

\noindent \textbf{Challenge.}  The adversary submits two equal length messages $M_0$ and $M_1$ and an access structure $\mathbb{A}^*$.  The adversary also supplies a new revocation list $RL^*$.
$RL^*$ and $\mathbb{A}^*$ satisfy the constraint that, for each user $u_k$, either $u_k \in RL^*$ or
$S_k$ does not satisfy $\mathbb{A}^*$.

The challenger flips a coin to obtain a random bit $b$ and returns $M_b$ encrypted with the access structure $\mathbb{A}^*$.  Additionally, it runs \textsc{ProxyRekey}$(PK, MK, RL)$ and returns the resulting key $PXK$ to the adversary.

\noindent \textbf{Phase 2.}  The adversary makes repeated queries to \textsc{Keygen} to obtain
keys for users $u_{q_1+1}, \ldots, u_{q_2}$ with attribute sets $S_{q_1+1}, \ldots, S_{q_2}$.  The new keys have to satisfy that if $u_k \notin RL^*$, then $S_k$ does not satisfy $\mathbb{A}^*$.

\noindent \textbf{Guess.}  The adversary outputs a guess $b'$ of $b$.

The advantage of an adversary is defined as $Pr[b' = b] - \frac{1}{2}$.

\newtheorem{defn}{Definition}
\begin{defn}
	A ciphertext-policy attribute-based encryption with proxy revocation scheme is secure if all polynomial time adversaries have at most negligible advantage in the above game.
\end{defn}

\newtheorem{theorem}{Theorem}

\subsection{Proof Sketch}

We can prove the security of our scheme using a variant of a generic bilinear group model.  Note
that since the security of the original CP-ABE scheme relies on the generic bilinear group model, the assumption we make is only slightly stronger than the original.  In particular, we must work within an asymmetric bilinear group, with a pairing of $e: \mathbb{G}_1 \times \mathbb{G}_2 \rightarrow \mathbb{G}_T$, such that there is no efficiently computable isomorphism from $\mathbb{G}_1$ to $\gg_2$.  (In a symmetric bilinear group, a user could submit $D_j'$ to \textsc{Convert} and recover $g_0^{r_j P(0)}$, obviating the need to use the proxy in further decryptions.)  This is believed to hold true for MNT curves~\cite{mnt}.

\noindent \textbf{The generic asymmetric bilinear group model.}  Consider three random
encodings of the additive group $\ff_p$ represented by injective maps $\psi_1, \psi_2, \psi_T : \ff_p \rightarrow \{ 0,1 \}^m,$ where $m > 3 \log p$.  We will define $\gg_1, \gg_2, \gg_T$ as the range of the respective map.  We are given access to a group action oracle for each group  and an oracle for a non-degenerate bilinear map $e: \gg_1 \times \gg_2 \rightarrow \gg_T$ (we will refer to the ranges of $\psi_1, \psi_2, \psi_T$ as $\gg_1, \gg_2, \gg_T$, respectively).  We are also given oracle access to the isomorphism $\phi: \gg_2 \rightarrow \gg_1$ and a hash function $H: \{ 0,1 \}^* \rightarrow \gg_2$.  Finally, we let $g_i = \psi_i(1)$ for $i=1,2$.

\begin{theorem}
	The construction presented in Section~\ref{cons} is secure under the generic asymmetric bilinear group model.
\end{theorem}

We sketch the main argument here; most of the rest of the details are similar to the proof presented by Bethencourt et al.~\cite{cpabe}.

\begin{proof}[Sketch]
First of all, we can assume that no ``unexpected collisions'' happen between the maps, meaning that if we keep track of the algebraic expressions passed to $\psi_1, \psi_2, \psi_T,$ and $\phi$, two values are equal if and only if the expressions are symbolically equivalent.  This assumption is true except for a negligible probability.

For simplicity of presentation, we will assume that $\aa^*$ contains a single attribute $A_j$ for some $j$.  Then after phase 2, the adversary has the following elements available:

$\gg_1: g_1, g_1^\beta, C = g_1^{\beta s}, C_j = g_1^{s} (= g_1^{q_j(0)}), C_j' g_1^{h_j s}$,

\noindent where $s$ is the random secret used to encrypt the challenge message and $h_j$ is implicitly defined such that $H(j) = g_2^{h_j}$.
\begin{align*}
\gg_2: g_2, D = g_2^{(\alpha + r_{u_k})/\beta}, D_j = g_2^{r_u + h_j r_{u_k,j} P(0)}, \\
D_j = g_2^{r_{u_k,j}}, D_j'' = g_2^{r_{u_k,j} P(u_k)}
\end{align*}

\noindent for each queried user $u_k$ where $A_j \in S_k$.  Note that we can ignore all $D_{j'}$ for $j' \neq j$ because, as with CP-ABE, they will not help with decryption.

\[ \gg_T: e(g_1,g_2)^\alpha, M \cdot e(g_1,g_2)^{\alpha s} \]

In addition, the adversary knows $u_k, P(u_k)$ for all the revoked users in $RL^*$.
Note that we can ignore any other elements of $\gg_1$ obtained through calls to \textsc{Convert} during Phase 1, or through calls to the isomorphism.  This is because in order to guess correctly with a non-negligible probability, the adversary needs to compute $e(g_1,g_2)^{\alpha s}$.  Since there are no occurrences of $s$ in $\gg_2$, each pairing must involve $g_1^{s k}$ for some $k$, and hence be derived from $C, C_j$, or $C_j'$.

The adversary can compute $e(C,D^{(u)}) = e(g_1,g_2)^{\alpha s + r_u s}$, hence computing $e(g_1,g_2)^{\alpha s}$ is equivalent to computing $e(g_1,g_2)^{r_u s}$ for some user $u$.  Note also that the secret keys obtained for other users are \emph{not} helpful here, for the same reason as in original CP-ABE. Algebraically, the adversary must solve the following equation:
\[ e(g_1,g_2)^{r_u s} = e\left(g_1^{x}, \left(D_j^{(u)}\right)^y \right) e(g_1,g_2)^z \]
\noindent where $x,y,$ and $z$ are derived from the available elements \emph{other} than
$D_j^{(u)}$.  Note that $x,y \neq 0$, since otherwise there is no way to introduce $r_u$ into the right-hand side.  However, the rest of the values are \emph{independent} of $P(0)$, since the only values of the polynomial available to the adversary outside of $D_j^{(u)}$ are $P(u_k)$
for $u_k \in RL^{*}$, and thus are not sufficient to determine $P(0)$.

\end{proof}

\section{Case Study: Social Networks}\label{arch}

Online Social Networks (OSNs) such as Facebook, Google+, Twitter, and LinkedIn are becoming
one of the most popular ways for users to interact online.  Besides personal
communication, OSNs provide the perfect platform for online games and other
applications. Users share personal information with the social network provider,
and trust the provider to protect their sensitive information.
However, this introduces privacy risks, as the collection of information is an attractive attack
target~\cite{phishing}.  Insiders can also release private information either intentionally or accidentally~\cite{flickr,gross}.  Several recent privacy compromises have thrown these issues into sharp focus~\cite{facebook-email-snafu,fb-article} .

These issues have motivated researchers to consider a paradigm shift, where instead of
trusting social network operators and being dependent on them to enforce
privacy, \textit{users} are in control of who views their data, for example, via
encryption~\cite{persona, lucas:wpes08, guha:wosn08, luo+:passat09}.
Fine-grained access control is a key challenge in this space; for example,
Facebook and LiveJournal have rolled out mechanisms to specify access control policies for each post, as the data items are usually destined for a subset of friends, or groups.

Persona~\cite{persona} is a state-of-the-art design that proposes the use of ABE to enable fine-grained access control. In OSNs, ABE allows users to have complete control over who can see their data, free from the whims of the OSN provider.  A user can create groups by assigning different attributes and keys to her social contacts, and then encrypt data such that only particular users having the desired set of attributes can decrypt it. This provides information protection from unauthorized users on the OSN, third-party application developers, and above all
the OSN provider itself.

However, groups are dynamic and therefore user attributes may change over time.
This could be because of change in location, work environment, or the nature or
strength of the relationship with a contact.  Recent studies have
shown that the user interaction graph is much less dense than friendship graph~\cite{interaction},
indicating that users interact most frequently with a small group of friends,
further validating the need for fine-grained access control. Moreover, the churn
rate for the interaction graph has been shown to be quite
high~\cite{interaction}, motivating the need for access control mechanisms to
support \textit{dynamic groups}.

Persona and similar designs introduce significant overhead for group membership changes, especially when a contact is removed from a group: all other members of the group must receive a new key; additionally, all existing data items destined for that group must be re-encrypted.  This does not scale when group sizes are large and group churn rate is high.

\begin{figure}[htbp]
\begin{center}
\includegraphics[width=0.8\columnwidth]{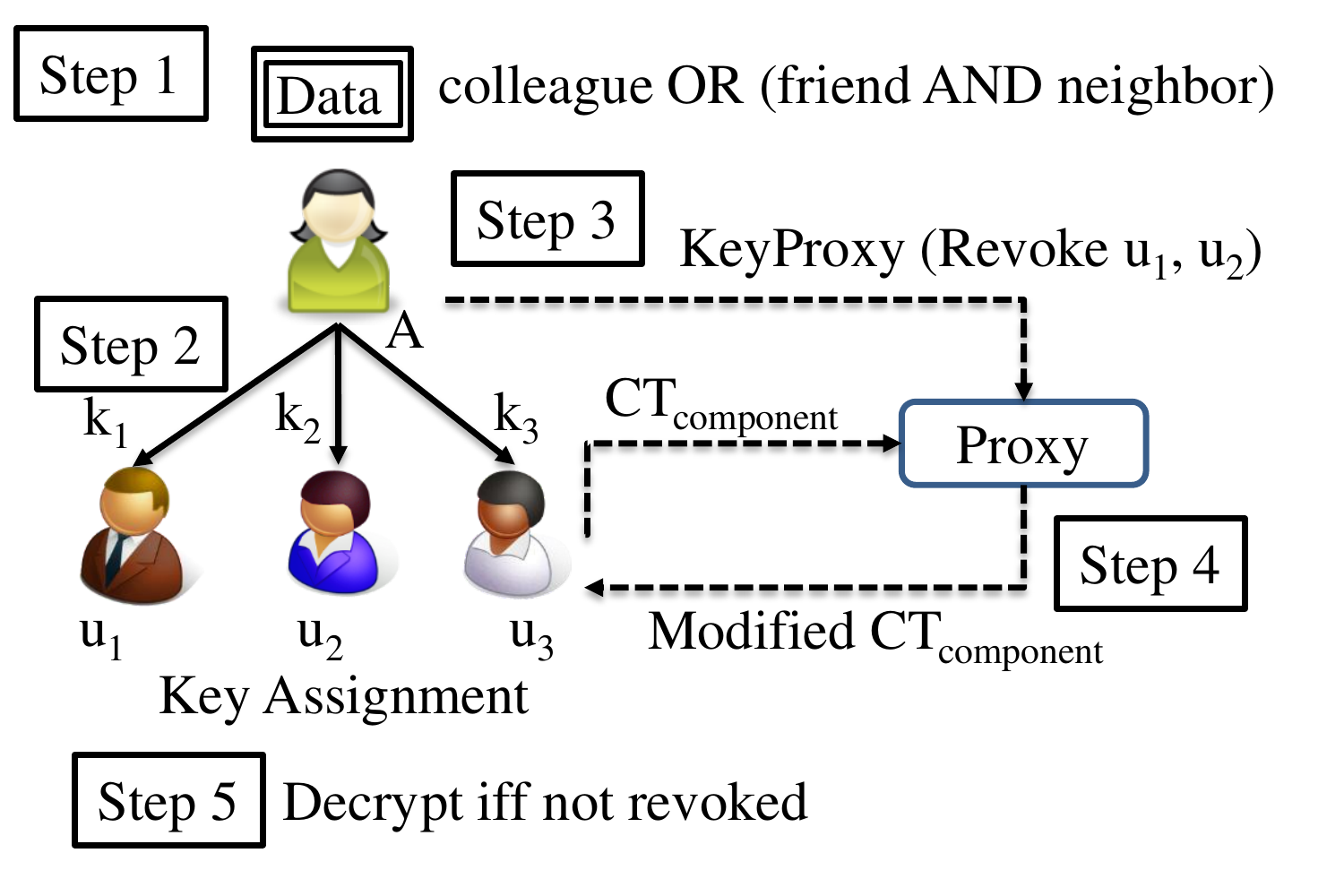}
\caption{Architecture of \easier -- OSN Built Using \rabe}
\label{fig:arch}
\end{center}
\end{figure}

We propose an architecture for OSN using \rabe as the underlying cryptographic scheme. Encryption successfully hides information from unintended parties. Traditional OSNs allow people to establish only a common relationship with each other by adding them as friends. Further specialization is achieved by creating lists and adding friends to that list. Relationships based on attributes make this task more expressive. 

Figure~\ref{fig:arch} shows the architectural overview of an OSN built using \rabe. We call this architecture \easier~\cite{easier}. Users in \easier become their own key authorities, define relationships by assigning attributes and relevant keys to the parties, and encrypting data under a policy that, if satisfied, allows decryption to the intended parties. For instance, user A assigns keys for attributes (\emph{colleague, neighbor}) to user B and encrypts data for the policy `\emph{colleague} OR \emph{(friend} AND \emph{neighbor)}'.

As mentioned earlier, the user interaction graph is different from the
friendship graph.  People interact with a subset of their defined contacts most
of the time. Recent privacy setting changes in wall posts in Facebook also
supports this fact. This requirement needs ABE in OSNs to consider revocation of access without re-encrypting all the data and re-keying everyone in the group when
access is denied to a contact. For example, after encrypting data under the mentioned
policy, user A might want everyone except $u_{1}$ and $u_{2}$ in her
\emph{colleague} group to decrypt the ciphertext, either temporarily or permanently.
Besides, A might want to revoke the \emph{colleague} attribute from
some of her contacts. \rabe provides this option by introducing a minimally
trusted proxy.

Upon revocation, the owner supplies enough information, in this case, a set of
$t$ (the maximum number of revoked users allowed) revoked users to the proxy to
construct a proxy key. The proxy is minimally trusted. Hence, the OSN provider
or a third party can act as proxy. An unrevoked user sends a specific set of
components of the ciphertext (details described earlier) to the proxy. The proxy
uses its key to change these components such that only unrevoked users can use
it, mathematically combine it with the rest of the ciphertext and their keys,
and finally obtain the plaintext. Since the conversion involves only the
lightweight components of ciphertext, it is not expensive, as we will
demonstrate later through experiments.


However, \rabe does not allow the proxy to decrypt the data since it does not
have the attribute keys. A new proxy key, created each time a revocation takes
place, prevents revoked users to collude with the proxy or with each other to
get the data. This prevents the revoked users from decrypting even old data
unless they store it somewhere. We argue that this is a desirable property: currently trusted contacts are not likely to crawl the entire set of social network data and store it for later use, but former friends or colleagues might try to abuse their former status by accessing past data.

Another example of an OSN that utilizes \rabe is DECENT, a decentralized architecture for OSN~\cite{decent-sesoc2012}. It uses \rabe for cryptographic protection, DHT for efficient data storage, and an object oriented architecture for flexible data representation.

\vspace{-15pt}
\subsection*{Friend-of-friend:}
A challenge in using cryptography to enforce access control in OSN is to support degrees of relationships, such as \textit{friend-of-friend} or \textit{contact-of-contact} to be more generic. \easier
 handles this challenge by using the feature attribute delegation in the distributed setting described before. 

In this approach, a user \ba generates a secret key for her contact \bb with the
attributes she wants to assign to him. She also adds an attribute named \textit{fof} to the key.
\bb uses the access delegation algorithm to delegate this specific attribute
(fof) to his contacts, for example \bc. Whenever \ba encrypts a piece of data with the
policy \textit{attr1, attr2, \ldots OR fof}, \bc can decrypt it with the
delegated key that he received from \bb. As in distributed attribute delegation, if \ba revokes \bb , or \bb
revokes \bc, the decryption will not succeed since both the proxies are required to participate
in the decryption, and the delegated secret key $\tilde{SK}$ contains
information about \ba, \bb, and \bc.

\section{Implementation and Experiemntal Evaluation}\label{exp}

We implemented the constructions in \rabe, as described in Section~\ref{cons}.
Our implementation involves introducing new components as well as modifying
different parts of the BSW CP-ABE toolkit~\cite{cpabe-toolkit}. The current
implementation supports complete key revocation and access delegation in a
distributed setting. Similar techniques can be applied to modify it to perform
attribute revocation. 

The implementation uses MNT curves~\cite{mnt} with $159$ bit
base field. All the experiments were carried out on a 2.40\,GHz Intel Core 2
Duo, 4\,GB memory, and running Ubuntu 10.04. We also implemented a Facebook
application to provide the functionality on a social network. The code and the Facebook application are available at \texttt{\url{http://www.soniajahid.com}} and \texttt{\url{http://apps.facebook.com/myeasier}} respectively.

\subsection{Performance Analysis}
\begin{figure*}[htbp]
\begin{center}
\subfloat[Secret Key Generation]
{
	\includegraphics[width=0.3\textwidth]{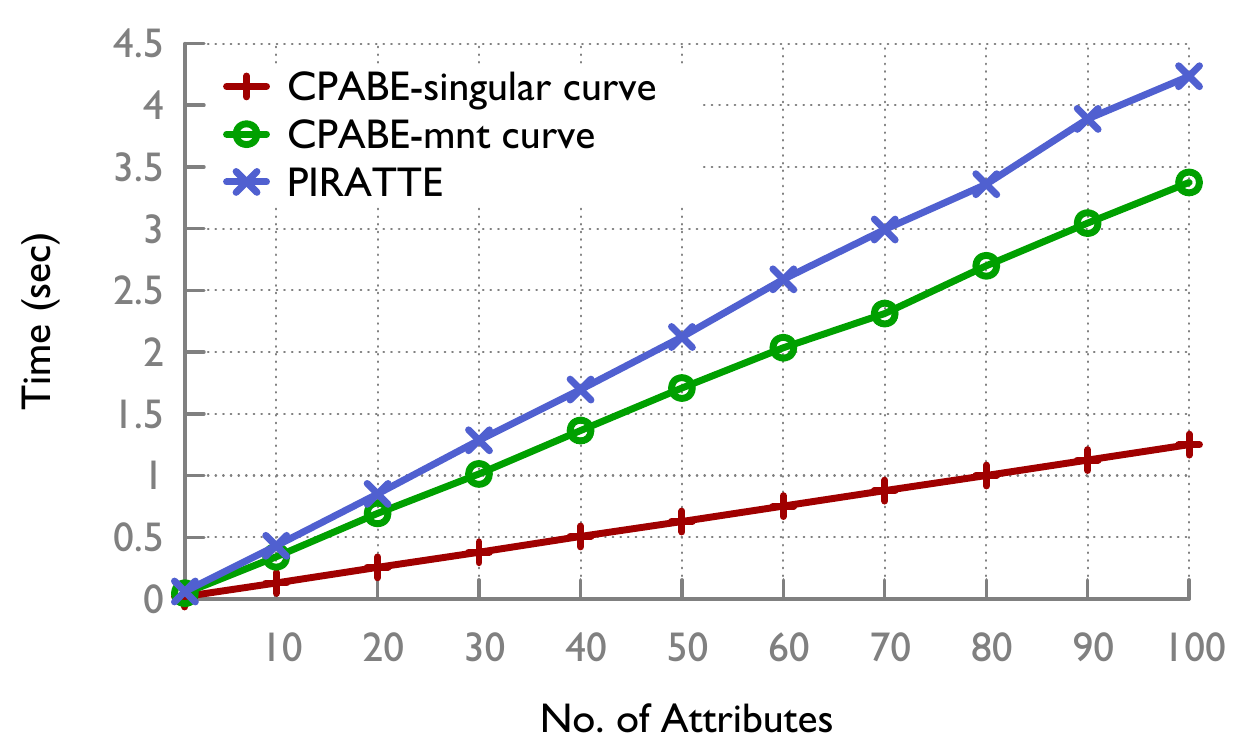}
	\label{chart:kgen}
}
\hfil
\subfloat[Encryption]
{
	\includegraphics[width=0.3\textwidth]{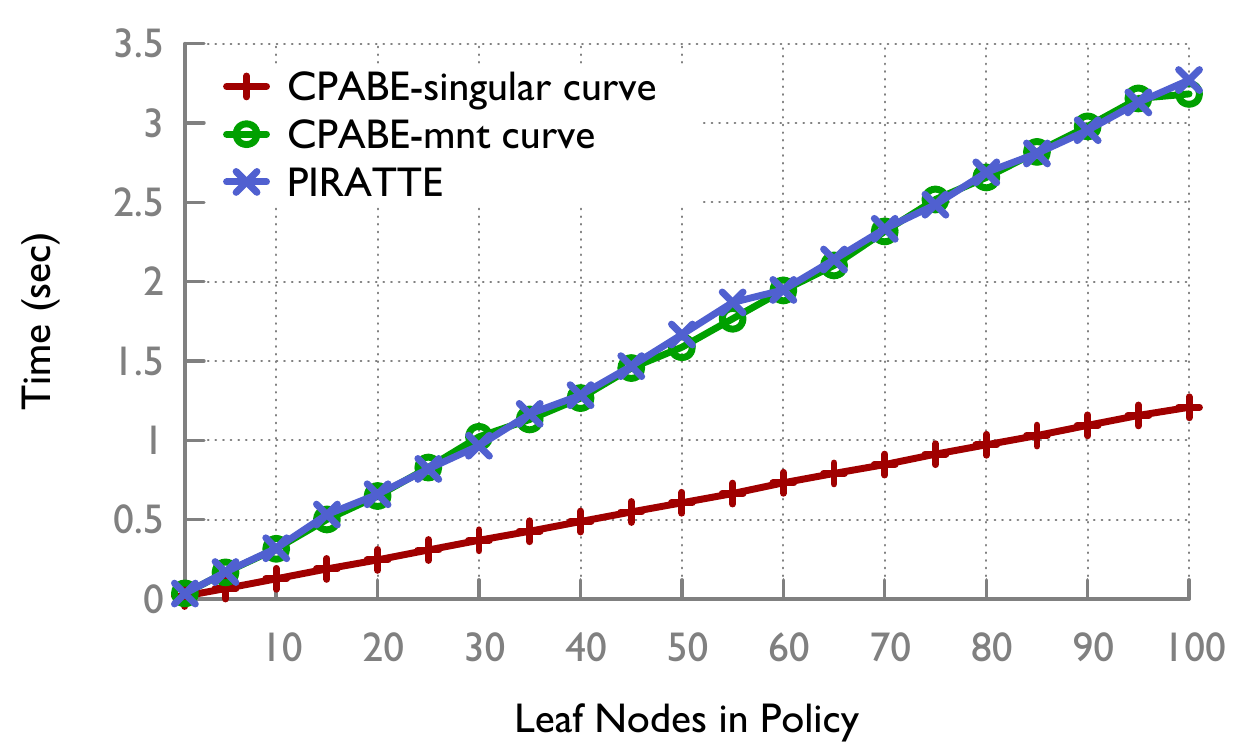}
	\label{chart:enc}
}
\hfil
\subfloat[Decryption]
{
	\includegraphics[width=0.3\textwidth]{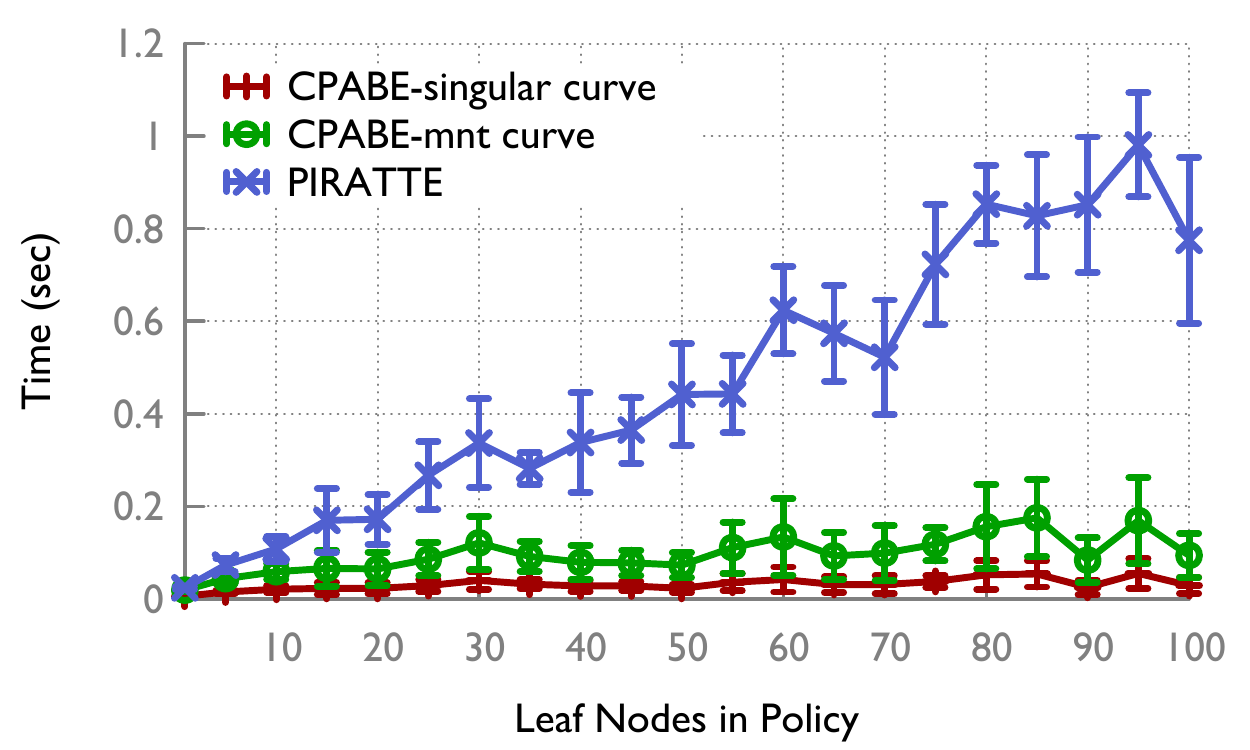}
	\label{chart:dec}
}
\vfil
\subfloat[Proxy Rekey in \rabe]
{
	\includegraphics[width=0.3\textwidth]{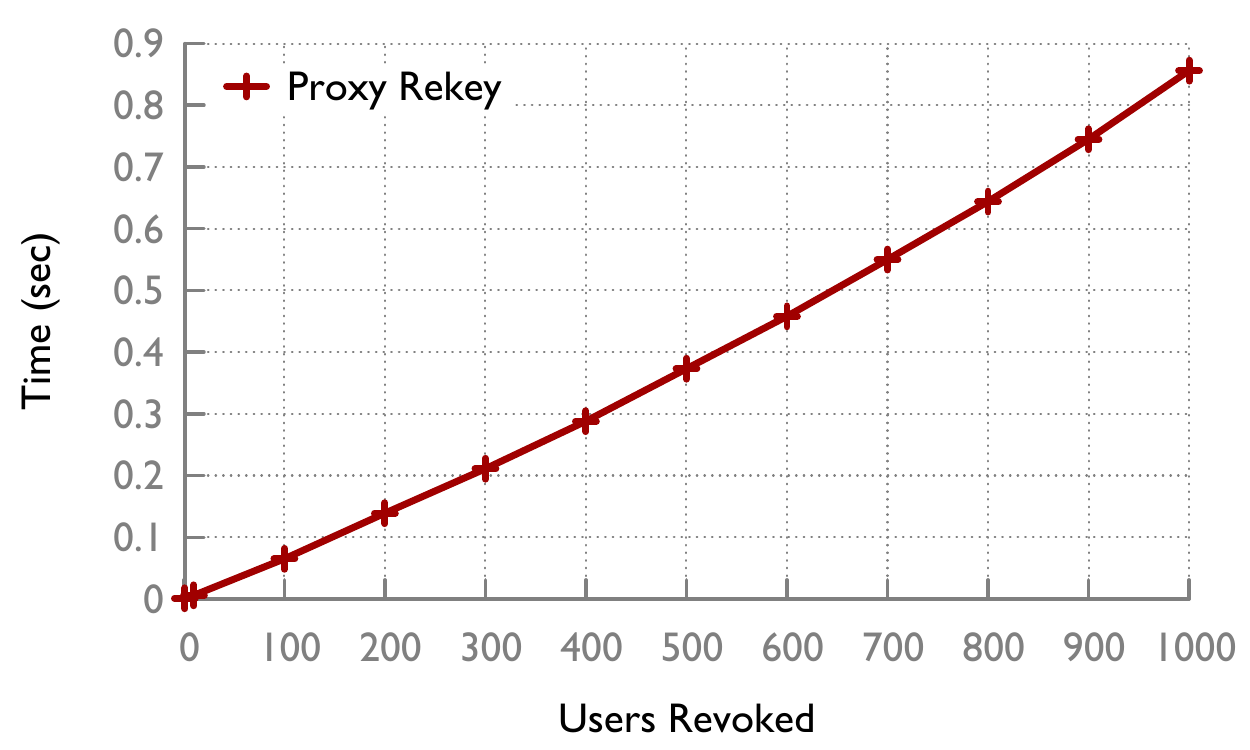}
	\label{chart:rvk}
}
\hfil
\subfloat[Conversion Pre-calculation by Proxy]
{
	\includegraphics[width=0.3\textwidth]{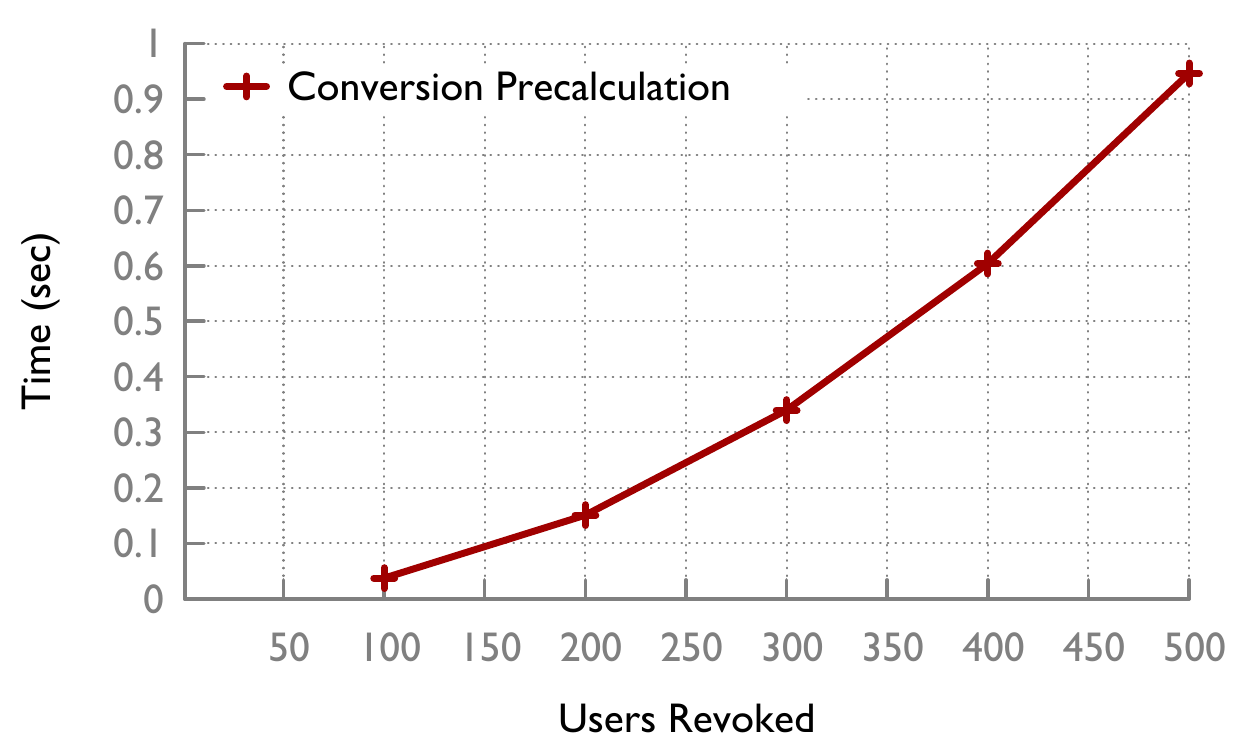}
	\label{chart:conv_precalc}
}
\hfil
\subfloat[Conversion in \rabe]
{
	\includegraphics[width=0.3\textwidth]{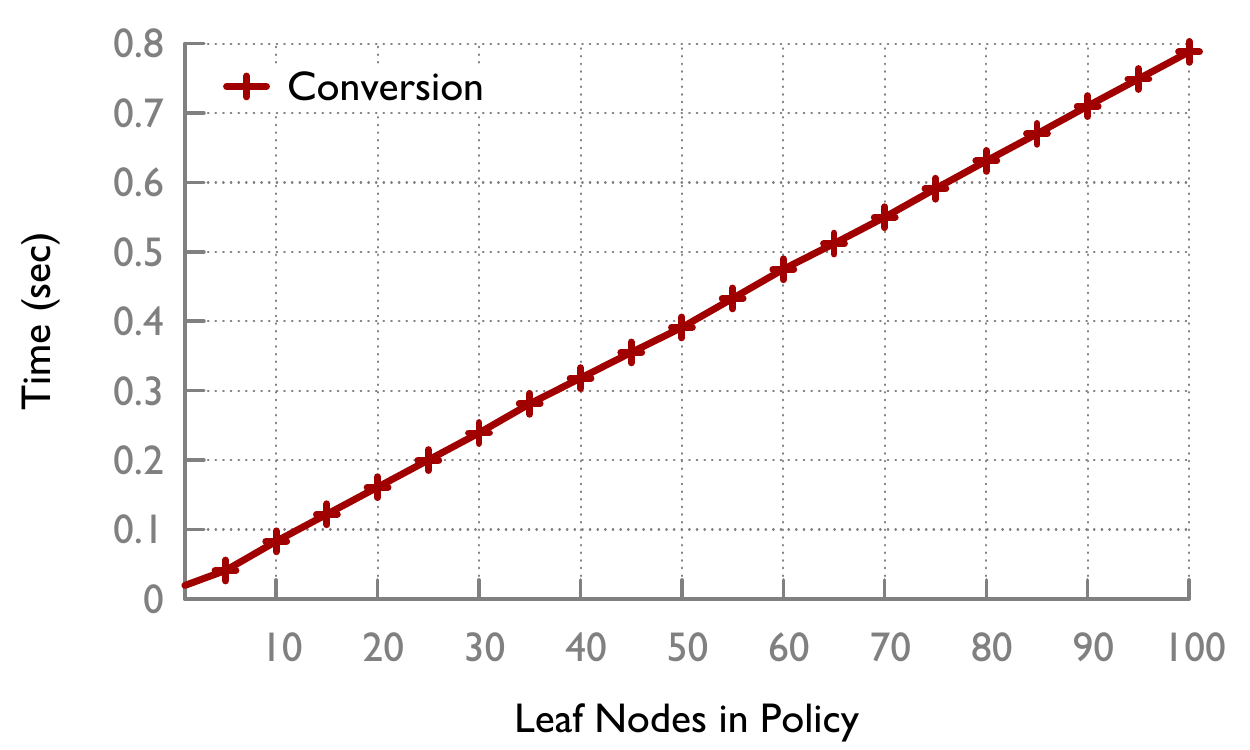}
	\label{chart:conv}
}

\caption{Performance Analysis of \rabe and Comparison with BSW CP-ABE}
\label{chart:time}
\end{center}
\end{figure*}

We provide some information on the performance evaluation of \rabe, and compare
it with CP-ABE both with MNT and super-singular curves. Though CP-ABE
implementation uses symmetric pairing, we use asymmetric pairing for both \rabe
and CP-ABE in our implementation. This provides security by preventing key and
ciphertext components exchange (discussed in Section~\ref{cons}). The results are shown in
Figure~\ref{chart:time}.

\textbf{Key Generation:} Key generation time is linear with number of attributes
both in CP-ABE and \rabe. Since it does an extra exponentiation, and generates
an extra component for each attribute in \rabe, the result is justified. CP-ABE
with supersingular curve requires the least time.

\textbf{Encryption:} To test encryption and decryption, we randomly generated
$10$ different policies for each of the desired number of leaves $(1, 5, 10,
\ldots 100)$. Encryption is also linear with respect to the number of leaf nodes
in the policy. We did not make any changes to the encryption scheme, so CP-ABE
with MNT and \rabe both take the same amount of time.

\textbf{Decryption:} Decryption depends on the policy used in encryption, and
the attributes involved. We generated a decryption key with $100$ attributes. It
has the superset of all the attributes used to generate the policies, and so
satisfies each of them. The decryption results are shown with a $95\%$
confidence interval. All the lines show the performance when an optimization was
used to ensure the usage of the minimum number of leaves in the algorithm
\textit{DecryptNode}. The required time is still below $1$ second.

\textbf{Proxy Rekey and Conversion:} \rabe involves two extra costs before
decryption: re-keying the proxy and converting the ciphertext components
specific to the leaves in the policy. The re-keying results
(Figure~\ref{chart:rvk}) show that for even $1000$ revoked users, the time
required is less than $0.9$ seconds. This should be compared with the time
required to rekey the rest of a group, i.e., generate a new key for everyone,
when even one person in the group is revoked.

Conversion primarily involves one exponentiation for each of the leaf specific
ciphertext components. It also calculates $\lambda_k$ for the requester $u_k$,
and completes the $\lambda_{i}s$ for each of the revoked users.  We perform an
optimization by allowing the proxy to pre-calculate a portion of the
$\lambda_i$'s in \texttt{Convert}. With the optimization, the proxy needs to do
$1$ multiplication per revoked user to calculate $\lambda_i$. It works as
follows:
\begin{align*}
&\lambda_{i}' = \prod_{u_i,u_j \in RL, i \neq j} \frac{u_{j}}{(u_{j}-u_{i})},
\text{ and }  l_{i}' = \lambda_{i}' P(u_{i})
\end{align*}

\begin{align*}
l_{i}  = l_{i}' \frac{u_{k}}{(u_{k}-u_{i})} = \lambda_{i}'
\frac{u_{k}}{(u_{k}-u_{i})} P(u_{i}) = \lambda_{i} P(u_{i}), \\
\forall \ u_i \in RL, u_{k} \not\in RL
\end{align*}

Figure~\ref{chart:conv_precalc} shows the time required for the proxy to do the
pre-calculation. Note that this is a one-time cost each time the proxy is
re-keyed and is not faced by users. Figure~\ref{chart:conv} shows the time
required to actually convert a ciphertext. The results are almost equal for the
number of revoked users since time to do $t$ exponentiation dominates the time
to do $t$ multiplication. Figure~\ref{chart:conv} shows the time for $500$
revoked users. We expect the proxy to be more powerful in terms of computing,
and hence rekeying, and conversion should be faster in practice. A user
performing decryption only faces the conversion time shown in
Figure~\ref{chart:conv} along with the decryption time mentioned earlier.

\textbf{Decryption with Delegated Key:} We measured the time required to decrypt
a ciphertext with delegated secret key in the multiple authority setting since we used it in \easier. We are interested in just one attribute,
i.e., \textit{fof} being delegated for the OSN setting. The policies used to
encrypt the data contain the \textit{fof} attribute in the form: \textit{(attr1,
$\ldots$) OR fof}. Therefore, the decryption needs to verify one attribute in
the policy. Hence, the time required does not depend on the number of leaf nodes
in the policy, and is constant. In our experiment, this time is about $0.34$sec.

\begin{table}[htbp]
\scriptsize
    \caption{Element Size}
    \begin{center}
    \begin{tabular}{|c|c|}
    \hline
    \bf Group & \bf Size (bytes) \\
    \hline
    $\gg_1$ & $44$\\ \hline
    $\gg_2$ & $124$\\ \hline
    $\gg_T$ & $124$\\ \hline
    $Z_p$   & $24$\\ \hline
    \end{tabular}
  \end{center}
  \label{tab:elemsize}
\end{table}

\vspace{-10pt}
 \begin{table}[htbp]
\scriptsize
    \caption{Component Size}
  \begin{center}
    \begin{tabular}{|c|p{1.1in}|p{1.1in}|}
    \hline
    \bf Component & \bf \rabe (bytes) & \bf CP-ABE (bytes)\\
    \hline
    Public Key & $1316$ & $1316$\\ \hline
    Master Key & $152 + (t+1)24$ & $148$\\ \hline
    Private Key & $128 + (a + 212) n$ & $128 + (a + 168)n$\\ \hline
    Ciphertext & $168 + 8i + (176+a)l$ & $168 + 8i + (176+a)l$\\ \hline
    Proxy Key & $24t$ & NA\\ \hline
    $C_{y}''$ & $124 l$ & NA\\ \hline
    \end{tabular}
  \end{center}
\vspace{-10pt}
    \label{tab:size}
\end{table}

\textbf{Component Size and Communication Overhead: } Table~\ref{tab:size} shows
the sizes of the components involved in the system for complete key revocation.
Size of the components for attribute revocation can be calculated similarly.
Elements from $\gg_1, \gg_2, \gg_T$, and $Z_p$ require $44, 124, 124$, and $24$
bytes respectively to represent(Table~\ref{tab:elemsize}). Users have to communicate with the proxy for
conversion by sending $C_{y}'$, and receiving $C_{y}''$. These are represented
using elements from $\gg_2$. This requires $124$ bytes to represent ($120$ for
the actual data, and $4$ for the variable size). Hence, conversion of a
ciphertext with $l$ leaf nodes in the policy will need to transfer $124l$ bytes
each way. The user also sends $u_k$, and receives $\lambda_k$ back. These are
represented using $Z_p$ which requires $24$ bytes. 

Public Key consists of a string describing the pairing used ($980$ bytes), $g_1$
and $h$ from $\gg_1$, $g_2$ from $\gg_2$, and $e(g_1, g_2)^\alpha$ from $\gg_T$.
Master Key consists of $\beta$ from $Z_p$, and ${g_2}^\alpha$ from $\gg_2$ in
CP-ABE. In \rabe, it also consists of a polynomial of degree $t$. The polynomial
consists of an integer $t$, and $t+1$ coefficients from $Z_p$. Private Key
consists of $D$ from $\gg_2$, number of attributes $n$ (integer), and $n$ of
$\langle D_j, D_j' \rangle$s from $\gg_2$ and $\gg_1$ respectively and
attributes of length $a$ (string) . \rabe also contains $n$ of $D_j''$ from
$\gg_1$. Ciphertext consists of $\tilde{C}$ from $\gg_T$, $C$ from $\gg_1$, and
components for each node in the policy. Both intermediate ($i$ nodes) and leaf
($l$ nodes) nodes have a threshold $k$ (integer), and number of children ($0$
for leaf, also an integer). A leaf node has a string attribute of length $a$,
and $C_y$ and $C_y'$ from $\gg_1$ and $\gg_2$ respectively. Proxy Key consists
of $t$ $Z_p$ elements.

\textbf{Attribute Revocation:} We can estimate the time required to perform
attribute revocation. All the algorithms work similarly. The only extra work is
in \texttt{Setup} where instead of just one polynomial, a polynomial is
generated for each attribute in the system. It requires $0.08$sec to generate a
polynomial of degree $100$ and increases linearly with the degree of the
polynomial. Hence, generating a polynomial for each attribute is scalable.

\subsection{Facebook Application}
Finally, we developed a Facebook application for \rabe as a proof of concept.
The goal is to present a high level idea of how an OSN that uses \rabe will
look like. We focused on Facebook because of lack of deployment of decentralized
architectures like Persona and Diaspora~\cite{diaspora}. In the current
implementation, all protocol functions are performed at our application server.
Moreover for convenience, we chose the client's proxy server to be the
application server itself.  

\begin{figure}[htbp]
\begin{center}
\subfloat[Main Page]
{
	\includegraphics[width=0.9\columnwidth]{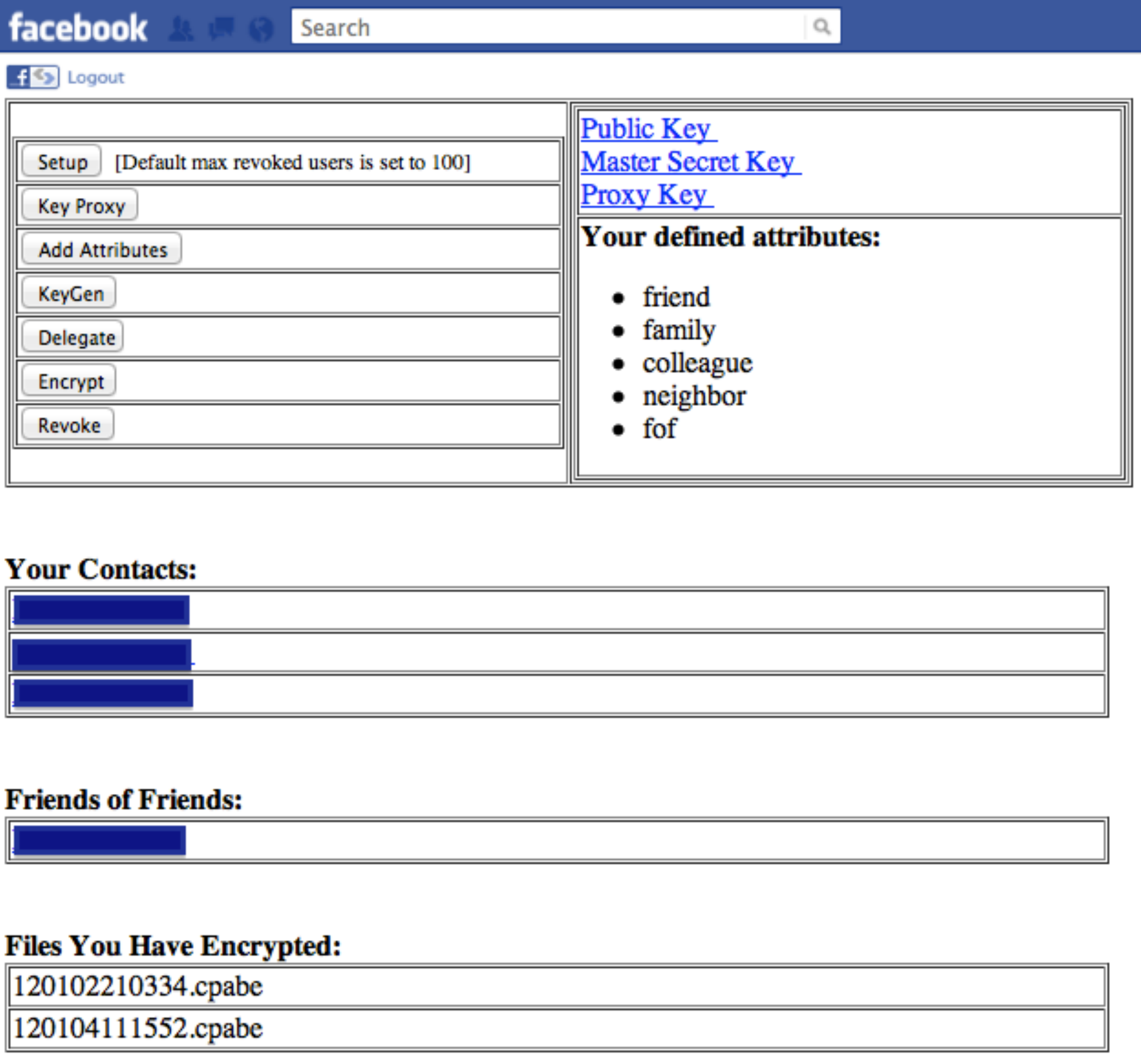}
	\label{fig:myeasier}
}
\vfil
\subfloat[Revoke]
{
	\includegraphics[width=0.9\columnwidth]{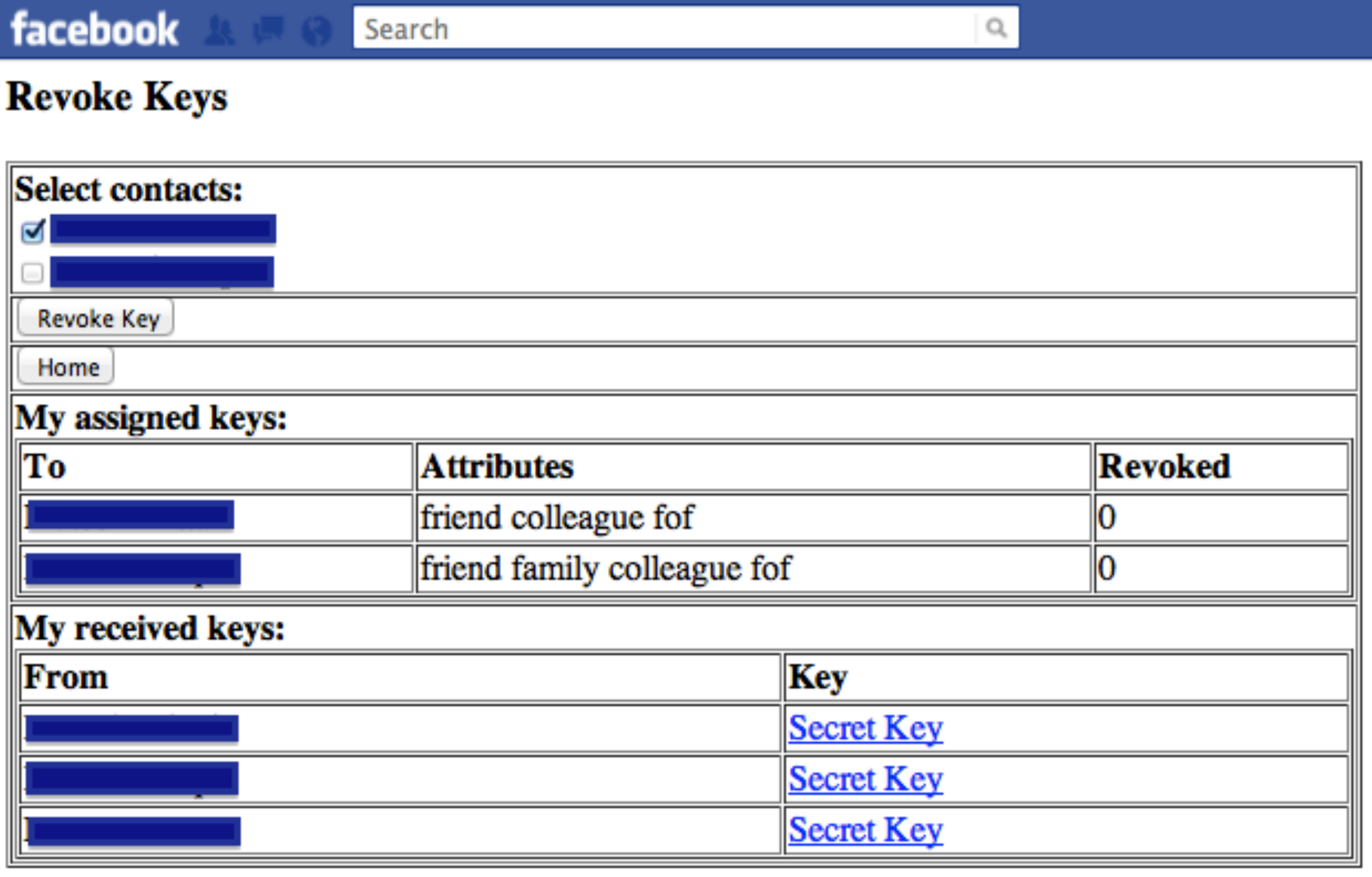}
	\label{fig:myeasier-rvk}
}
\vfil
\caption{\rabe on Facebook}
\label{fig:fb-app}
\end{center}
\end{figure}

The application retrieves public profile information of the users installing it
using Facebook API, and uses this data as its user profile information.
Figures~\ref{fig:myeasier} and~\ref{fig:myeasier-rvk} depict screen-shots of the
same. Data is hidden for privacy purposes.

When a user first installs our Facebook application, an account is created in
the application server. We provide a brief description of the supported
functionality.
\squishlist
\item \textbf{Setup}: The \emph{Setup} button is used to initialize a public key PK and a master secret key MK.
\item \textbf{Key Proxy}: The \emph{Key Proxy} button is used to generate a key for the
proxy server. This is used to key the proxy when there is no revocation. Each
revocation updates the proxy key, so there is no need to manually rekey the
proxy with each revocation.
    
\item \textbf{Add Attributes}: The \emph{Add Attributes} functionality is invoked to
define a set of attributes like \emph{family, coworker, researcher}, etc.

\item \textbf{KeyGen}: The \emph{KeyGen} functionality is used to generate keys for
Facebook friends who also have the \rabe application installed. These users are assigned
specific attributes, with the selection of attributes being made from amongst
the choices defined in the Add Attributes step.

\item \textbf{Encrypt}: The \emph{Encrypt} button is used to generate a ciphertext
under a policy defined over the available set of attributes. For convenience,
application users can choose and encrypt userid, about, and gender for their
contacts which is already retrieved from their profile information (if
available).

\item \textbf{Revoke}: Users can select one or more contacts to whom they
assigned keys
previously, and revoke the keys from them.

\item \textbf{Decrypt}: A user can click on any ciphertext generated by a
contact who assigned her a secret key, and is able to view the plaintext if his/her
attributes satisfy the ciphertext-policy. When a user clicks Decrypt, the
ciphertext is converted by the proxy, and then the user secret key is used to
decrypt the data. Names are  not shown in the snapshot because of privacy.

\item \textbf{Delegate}: The Delegate option is used to delegate the
\textit{fof} attribute from the keys a user received to her contacts. It shows
the list of contacts from whom a user received secret attribute keys and to
whom she can delegate access to. Duplicates and self-assignment are prohibited
through checks.

\squishend
\vspace{-20pt}
\section{Related Work}
\label{sec:related}

\subsection{Revocation Schemes}

Attrapadung and Imai address the revocation problem in~\cite{conj-revoc-abe} by combining broadcast encryption scheme with both CP-ABE and KP-ABE. This scheme requires knowledge about the list of all possible users during encryption, i.e., attaches one component per user to the ciphertext. Knowing the list of all possible users in advance is inconvenient for most scenarios, specifically OSNs. Bethencourt et al. in~\cite{cpabe} and Boldyreva et al. in~\cite{kpabe-revoc} propose expiration-based revocation scheme for CP-ABE and KP-ABE respectively. In these schemes, the secret keys (CP-ABE) or the ciphertext (KP-ABE) contain expiry time as an extra attribute. Time-based revocation may not be a desired property in several applications where an immediate revocation is necessary. It introduces a window of vulnerability, i.e., the gap between the desired time from revocation to the actual time of revocation. Ostrovsky et al. ~\cite{ostrovsky07nonmonotone} present a new KP-ABE scheme with non-monotonic access structure to support negation of attributes. Revocation can be implemented by adding a NOT-attribute to the policy in private key. However, this will require re-keying the users from whom an attribute is revoked.

Lewko et al. in~\cite{revoc-smallkey} propose a revocation scheme with small private keys. However, their approach also increases the size of the ciphertext by incorporating the list of revoked and unrevoked users with it. In~\cite{junbeom-cpabe-revoke1} Hur proposes a revocation scheme for CP-ABE where each leaf node, i.e. attribute in the policy used to encrypt the data contains a group of users who possess that attribute. The scheme consists of a key generation center (KGC) and a data storing center (DSC). The DSC is equivalent to the proxy in our proposed scheme. However, the approach requires secret key update for all the users of an attribute group from which at least one user was revoked; this also includes the non-revoked users in that group.

The CCA secure construction of Yu et al.~\cite{yu:asiaccs10} involves updating the master key component of each attribute that has been revoked in the system. The public key components are then updated and data is encrypted with the new public key. Finally, proxy re-keys are generated that enable a proxy to update the user secret keys to the new version for all but the revoked user. Basically the re-keying burden in placed on the proxy, instead of users. We note that the existing data would need to be re-encrypted by the proxy; placing a significant burden on the system. While our scheme is only CPA secure under the generic group model (weaker security than Yu et al.), it does not require re-encryption of existing data.

\subsection{Proxy Based Re-encryption}

Our revocation techniques are based on the notion of proxy re-encryption. We will now briefly trace some developments in this field, and discuss why the state of the art techniques cannot be directly applied  to our setting.

Blaze et al.~\cite{blaze:eurocrypt98} introduced the notion of {\it proxy re-encryption}, in which a proxy could convert a ciphertext for Alice into a ciphertext for Bob, using a  specially generated proxy key. The holders of public-key pairs $A$ (Alice) and $B$ (Bob) create and publish a proxy key $\pi_{A \rightarrow B}$ such that $D(\pi(E(m, e_A), \pi_{A \rightarrow B}), d_B) = m$, where $E(m, e)$ is the public encryption function of message $m$ under encryption key $e$, $D(c, d)$ is the decryption function of ciphertext $c$ under decryption key $d$, $\pi(c, \pi_{A \rightarrow B})$ is the proxy function that converts ciphertext $c$ according to proxy key $\pi_{A \rightarrow B}$,  and $e_A$, $e_B$, $d_A$, $d_B$ are the public encryption and secret decryption component keys for key pairs $A$ and $B$, respectively. In the El-Gamal based construction proposed by Blaze et al., while the proxy cannot see the plaintext message $m$, it can collude with $B$ to recover the secret key for $A$. Moreover, the construction is {\it bidirectional}, and the  proxy key can be  used to convert ciphertext for Bob into a ciphertext for Alice as well.

Ateniese et al.~\cite{ateniese:ndss05} propose {\it unidirectional} protocols for proxy re-encryption based on bilinear maps, where a re-encryption key from $A$ to $B$ does not imply a re-encryption key from $B$ to $A$. Canetti and Hohenberger~\cite{canetti:ccs07} proposed the first CCA secure bidirectional proxy re-encryption scheme, while Libert and Vergnaud~\cite{libert:pkc08} proposed the first CCA secure unidirectional proxy re-encryption scheme in the standard model. We note that all of the above schemes are limited to the public key setting. Green and Ateniese~\cite{green:acns07} extended the model to the identity based setting by proposing a scheme for identity based proxy re-encryption, but it was not until Liang et al.~\cite{liang:asiaccs09} that the attribute based encryption setting was considered. 

In the attribute based setting of Liang et al., a user could designate a proxy, who can re-encrypt a ciphertext with a certain access policy into another ciphertext with a different access policy. Furthermore, the authors showed their scheme to be selective-structure chosen plaintext secure. However, it is not possible to apply their construction to the problem of attribute revocation because their construction does not support negative attributes. 


\subsection{Social Network Privacy Architectures}

We will now describe existing social network privacy architectures and provide a comparison with \easier. We can categorize the different schemes based on the level of trust for centralized services: a) trusted central servers, b) untrusted central servers, and c) decentralized architecture. Intuitively, decentralized architectures provide the highest level of privacy. \easier is close to Persona~\cite{persona}, which is the state-of-the-art decentralized architecture for social network privacy. In almost all of the settings, access control is performed via encryption. We believe that the public key encryption setting is not well suited for fine-grained access control. Instead, \rabe uses attribute based encryption techniques to enable users to perform fine-grained access control. Moreover, none of the schemes focus on the issue of efficient user/attribute revocation.

\textbf{Trusted Centralized Architectures:}
Lucas and Borisov~\cite{lucas:wpes08} propose flyByNight, a facebook application designed to mitigate privacy risks in social networks. flyByNight users encrypt sensitive messages using JavaScript on the client side and send the ciphertext to some intended party, i.e., Facebook friends, who can then decrypt the data. The architecture ensures that transferred sensitive data cannot be viewed by the Facebook servers in an unencrypted form. However, the utility of flyByNight is limited to preserving the privacy of messages intended for social network friends, i.e., email type communication, and thus, it does not provide complete privacy. For example, the application server knows a user's friendlist on facebook. flyByNight is also vulnerable to active attacks by the OSN provider, since the OSN interface is used for key management.

Singh et al.~\cite{singh:usenix09} propose the xBook framework for building privacy preserving social
network applications. xBook uses information flow models to control what untrusted applications can do
with the information they receive. Their design retains the functionality offered by existing online social
networks. xBook provides enforcement for both user-user access control for data flowing within a single
application, as well as for information sharing with entities outside xBook. Social network applications are
re-designed to have access to all the data that they require, but this data is not allowed to be passed on to
an external entity unless approved by the user.

\textbf{Untrusted Centralized Architectures:}
Guha et al.~\cite{guha:wosn08} propose to improve user privacy while still preserving the functionality of existing online social network providers. Their architecture is called {\it None of Your Business} (NOYB), in which encryption is used to hide the data from the untrusted social network provider. The key feature of their architecture is a general cipher and encoding scheme that preserves the semantic properties of data such that it can be processed by the social network provider {\it oblivious} to encryption.  A user's private information is partitioned into {\it atoms}, and NOYB encrypts a user's atom by substituting it with the atom of another user. Thus the OSN can operate on ciphered data, but only the authorized users can decrypt the result.

Anderson et al.~\cite{anderson:wosn09} propose a client-server architecture for providing social network privacy. In their design, the server is a very simple untrusted social network provider which serves as a data container, while a complex client side architecture performs the access control. The server only provides availability and client is responsible for data confidentiality and integrity. In addition to content data, the architecture is also able to protect the social graph information.

\textbf{Decentralized Architectures:}
As described earlier, Persona~\cite{persona} is the state-of-the-art decentralized architecture for social network privacy. \easier is similar to Persona as both use ABE, but \easier also provides an efficient mechanism for revocation, thereby avoiding the overhead of re-keying with group members as well as re-encryption of old data.

In the approach by Shakimov et al.~\cite{shakimov:wosn09}, users store their data in a Virtual Independent Server (VIS) owned by themselves. These VISs form an overlay network per OSN. The authors consider different types of decentralized OSNs, depending on where the VISs reside: a) cloud b) desktop , and  c) hybrid based, and compare their privacy, costs and availability. Diaspora~\cite{diaspora} is a decentralized OSN that users install on their own personal web servers. Backes et al.~\cite{Backes11} present a core API for social networking, which can also constitute a plug-in for distributed OSNs. They assume that the server is trusted with the data while implementing access control. Both of these approaches avoid encryption. 

PeerSon~\cite{peerson}, LotusNet~\cite{lotusnet} and Safebook~\cite{safebook}, three decentralized designs for social networking benefit from DHTs in their architecture. PeerSon and Safebook suggest access control through encryption, but they fall short in providing fine-grained policies comparing to ABE-based access control. Safebook is based on a peer-to-peer overlay network named \emph{Matryoshka}. The end-to-end privacy in Matryoshka is provided by leveraging existing hop-by-hop trust. In all of these schemes overhead of key revocation affects performance.

\section{Conclusion}\label{conclude}
We presented a scheme called \rabe for efficient revocation in \emph{Ciphertext Policy Attribute-based Encryption}. We achieved this revocation scheme by introducing a semi-trusted proxy, leveraging ideas from a group communication scheme, and combining it with ABE. We also presented an architecture named \easier for Online Social Networks (OSNs) that enforces access control through encryption using techniques in \rabe. Although we showed the use of \rabe in an OSN setting, it can be applied to any context where  ABE is used for data protection with dynamic group membership. We implemented the scheme and compared it with Bethencourt et al.'s CP-ABE. Our results show that \rabe is scalable in terms of computation and communication for OSNs; accordingly, we have built a prototype application in the Facebook OSN to provide such encryption.  

\ifCLASSOPTIONcompsoc
  \section*{Acknowledgments}
\else
  \section*{Acknowledgment}
\fi
The authors would like to thank Prateek Mittal.

      
\bibliographystyle{IEEEtran}
\bibliography{paper}



\end{document}